\documentclass[aps,prd,onecolumn,groupedaddress,nofootinbib,preprintnumbers]{revtex4-2}
\usepackage{amssymb,amsmath,amsfonts}
\usepackage{graphicx,color,bm,cancel,comment,ulem}
\usepackage{hyperref}

\allowdisplaybreaks[4]

\begin{document}

\tolerance=5000

\title{Future singularity in an anisotropic universe}

\preprint{KEK-TH-2632}
\preprint{KEK-Cosmo-0348}

\author{Taishi~Katsuragawa}
\email{taishi@ccnu.edu.cn}
\affiliation{
Institute of Astrophysics, Central China Normal University, Wuhan 430079, China
}

\author{Shin'ichi~Nojiri}
\email{nojiri@gravity.phys.nagoya-u.ac.jp}
\affiliation{
KEK Theory Center, Institute of Particle and Nuclear Studies, 
High Energy Accelerator Research Organization (KEK), Oho 1-1, Tsukuba, Ibaraki 305-0801, Japan
}
\affiliation{
Kobayashi-Maskawa Institute for the Origin of Particles and the Universe, 
Nagoya University, Nagoya 464-8602, Japan
}
\author{Sergei~D.~Odintsov}
\email{odintsov@ice.csic.es}
\affiliation{
ICREA, Passeig Lluis Companys, 23, 08010 Barcelona, Spain
}
\affiliation{
Institute of Space Sciences (ICE, CSIC) C. Can Magrans s/n, 08193 Barcelona, Spain
}

\begin{abstract}
We investigate future singularities originating from the anisotropy in the Universe. 
We formulate a new class of singularities in the homogeneous and anisotropic universe,
comparing them with the known singularities in the homogeneous and isotropic universe.
We also discuss the physical consequences of the new singularities.
Moreover, we develop a novel reconstruction method for the anisotropic universe by introducing four scalar fields to reconstruct cosmological models in which future singularities appear.
We present an explicit example where the anisotropy may grow in the future up to singularity.
\end{abstract}

\maketitle

\section{Introduction}
\label{introduction}

The concordance $\Lambda$ cold dark matter ($\Lambda$CDM) model, including the cosmological constant and cold dark matter, has been in good agreement with observational data.
However, for several problems that are difficult to explain in the $\Lambda$CDM model,
cosmological models that go beyond the standard model have been intensively investigated in the context of modifying Einstein's gravity, known as the modified gravity theory.
In the search for the beyond-$\Lambda$CDM model, 
modifications of the gravitational theory have provided a variety of cosmological models,
and cosmological observations have indeed constrained the gravitational theories.
However, it is also significant to examine the cosmological principle on which the $\Lambda$CDM model stands; that is, the universe is homogeneous and isotropic spacetime on large scales and written as the Friedmann-Lema\^{i}tre-Robertson-Walker (FLRW) metric as the zeroth order approximation.

Cosmic microwave background (CMB) and large-scale structure data have thoroughly tested the cosmological principle. 
Recently, strong evidence of a violation of the cosmological principle of isotropy has been reported~\cite{Fosalba:2020gls, Yeung:2022smn, Aluri:2022hzs}.
Although there has been much discussion about the origin of the anisotropy and its time evolution, the cosmological no-hair conjecture~\cite{Wald:1983ky} provides a strong prediction, independent of the details of the model, that the anisotropy will exponentially decrease once inflation occurs.
However, it is still possible to evade the cosmic no-hair conjecture, and the anisotropic inflation models~\cite{Kanno:2008gn, Himmetoglu:2008zp, Himmetoglu:2008hx, Watanabe:2009ct, Maleknejad:2012as, Nojiri:2022idp, Allahyari:2023kfm} suggest that the spontaneous rotational-symmetry breaking could occur during the inflation.
The generated cosmological anisotropy could help us understand the CMB anomalies, and it is being vigorously studied along with other cosmological anomalies, such as Hubble tension.

In addition to studying the origin of anisotropy and its effects in the early universe, 
It is also essential to study how anisotropy will evolve in the future.
It has already been suggested that the current universe contains a small amount of anisotropy, and due to new physics or unveiled mechanisms, the future universe may develop a larger amount of anisotropy. 
It is feasible to construct cosmological models with potentially increasing anisotropy and also significant to investigate what may happen in the future within such models.
For example, we allow finite anisotropy in all cosmic history. In that case, the anisotropy may grow or even show singular behaviors because three spatial directions can evolve differently and include singularities.
We will explore cosmological models based on finite anisotropy and search for possible new physics that lies therein.

In this paper, we investigate the general homogeneous and anisotropic universe, 
assuming that anisotropy generated by some mechanisms exists in the universe.
We mainly discuss the future singularities generated by the anisotropy.
It has been known that cosmological models generally encompass five types of finite-time singularities in the FLRW universe~\cite{Nojiri:2005sx, deHaro:2023lbq, Trivedi:2023zlf}.
In contrast to these singularities known in the homogeneous and isotropic universe,
this paper presents new types of finite-time singularities in the homogeneous and anisotropic universe.
We discuss the classification of the new singularities and their physical meanings due to the anisotropy by analogy with the known singularities in the FLRW case. 
To demonstrate the growing anisotropy and associated singularities, 
we construct a cosmological model that realizes the finite-time singularities, 
by developing a new cosmological reconstruction method.

This paper is organized as follows.
In Sec.~\ref{sec-classification-singularities}, we briefly review the finite-time singularities known in the FLRW universe.
In Sec.~\ref{anisotropic_spacetime}, we formulate the cosmological model with the broken rotational symmetry and classify the finite-time singularities due to the finite anisotropy.
Moreover, we show phenomena caused by these singularities using the geodesic deviation equation.
In Sec.~\ref{reconstruction}, we demonstrate the reconstruction of the cosmological models where the finite-time singularities appear.
As a specific example, we use  Einstein's gravity as a benchmark gravitational theory.


\section{Finite-time Singularities in FLRW universe}
\label{sec-classification-singularities}

We briefly review the finite-time singularities in the FLRW universe, homogeneous and isotropic spacetime, following from Refs.~\cite{Nojiri:2005sx, deHaro:2023lbq,Trivedi:2023zlf}. 
The line element of the FLRW universe is given by  
\begin{align}
\label{Eq: FLRWk}
    d s^2 
    = -dt^2 + \alpha^2(t) 
    \left[\frac{dr^2}{1-Kr^2} + r^2 \left(d\vartheta^2 + \sin^2 \vartheta d\phi^2\right) \right]
    \, ,
\end{align}
where $\alpha(t)$ is a scale factor, and $(t, r, \vartheta, \phi)$ are the comoving coordinates.
$K$ describes three different geometries for three distinct values, namely, spatially flat ($K =0$), closed ($K > 0$), and open ($K < 0$). 
We consider the spatially flat FLRW, $K=0$, where Eq.~\eqref{Eq: FLRWk} reduces to the following form, 
\begin{align}
\label{Eq: FLRWk0}
    d s^2 
    &= -dt^2 + \alpha^2(t) \sum_{i=1,2,3}\left( dx^i \right)^2
    \, .
\end{align}

In Einstein's gravity, the expansion of the FLRW universe with $K=0$ in Eq.~\eqref{Eq: FLRWk0} is described by 
the Friedmann equation and the Raychaudhuri equation, 
\begin{align}
H^2=\frac{\kappa^2}{3} \rho\, , \quad \dot{H}=-\frac{\kappa^2}{2} \left(\rho+p\right) \, .
\label{FReqs}
\end{align}
We denote the Hubble parameter by $H\equiv \frac{\dot\alpha}{\alpha}$, where the dot represents the derivative with respect to $t$,
and $\kappa^2 = 8 \pi G_N$ with Newton's gravitational constant $G_N$.
$p$ and $\rho$ are the pressure and the energy density of the matter contents in the universe. 
We also introduce the equation of state (EOS) as $p=w\rho$, where $w$ is the EOS parameter. 

Based on Eq.~\eqref{FReqs}, the types of future singularities appearing in various cosmological models 
are classified as follows:
when $t\rightarrow t_s$,
\begin{enumerate}
    \item Type I (big rip) singularity: 
    $\alpha\rightarrow \infty$, $\rho\rightarrow\infty$ and $|p|\rightarrow\infty$.
    \item Type II (sudden) singularity: 
    $\alpha \rightarrow const.$ and $\rho \rightarrow const.$, but $|p|\rightarrow\infty$.
    $\alpha$ and $\dot \alpha$ are finite, but $\ddot \alpha$ diverges. 
    \item Type III (big freeze) singularity: 
    $\alpha \rightarrow const.$, but $\rho\rightarrow\infty$ and $|p|\rightarrow\infty$. 
    $\alpha$ is finite, but $\dot \alpha$ diverges.
    \item Type IV (generalized sudden) singularity: 
    $\alpha \rightarrow const.$, $\rho \rightarrow const.$, and $|p| \rightarrow const.$, 
    but some higher derivatives of $H$ diverge.
    $\alpha$, $\dot\alpha$, and $\ddot\alpha$ are finite, but higher derivatives of $\alpha$ diverge.
    \item Type V ($w$) singularity: 
    $w \rightarrow \infty$, but $p$ and $\rho$ are finite.
    This type depends on the properties of the matter, but the behavior of $\alpha$ is identical to that in type II, 
    that is, $\alpha$ and $\dot\alpha$ are finite, but $\ddot\alpha$ diverges.
\end{enumerate}
Type I singularity was first introduced in~\cite{Caldwell:2003vq}, which appears in the universe 
filled by phantom fluid~\cite{Caldwell:1999ew}. 
Type II singularity was proposed in~\cite{Barrow:2004xh}. 
Type III and type IV singularities were obtained by complementing the type I and type II singularities in~\cite{Nojiri:2005sx} (for type III, see also \cite{Bouhmadi-Lopez:2006fwq,Trivedi:2022ngt}). 
Although type I-IV singularities have completely classified the singular behaviors of spacetime, 
in \cite{Dabrowski:2009kg}, the singular behavior of the EOS parameter $w$ was also considered. 

To illustrate what could happen near the singularities,
we consider the geodesic deviation equation:
\begin{align}
\label{geodesicdev}
    \frac{D^2 S^\mu}{d\tau^2} = R^\mu_{\ \nu\rho\sigma}T^\nu T^\rho S^\sigma
    \, .
\end{align}
Here, $\tau$, $S^\mu$, and $T^{\mu}$ present the proper time, deviation vector, and the tangent vector, respectively. 
In the FLRW spacetime \eqref{Eq: FLRWk0}, we may choose $T^0=1$, $T^i=0$.
Then, Eq.~\eqref{geodesicdev} is reduced into
\begin{align}
\label{geodesicdev1}
    \frac{D^2 S^i}{d\tau^2} = R^i_{\ 00j} S^j
    \, .
\end{align}
In the FLRW universe, we have 
\begin{align}
\label{Riemann}
    R^{i}_{\ 00j}
    = \left( \dot H + H^2 \right) \delta_{ij}
    \, ,
\end{align}
and Eq.~\eqref{geodesicdev1} gives 
\begin{align}
\label{geodesicdev2}
    \frac{D^2 S^i}{d\tau^2} = \left( \dot H + H^2 \right) S^i \, .
\end{align}
$H$ and $\dot H$ diverge in type I and III singularities, and $\dot H$ diverges in type II singularity. 
Thus, Eq.~\eqref{geodesicdev2} tells us that spacetime is ripped. 
There could be the case that $H$, $\dot H$, or both go to infinity in the infinite future. 
Even in this case, everything is ripped finally, which is called a little rip~\cite{Frampton:2011sp, Brevik:2011mm, Frampton:2011rh}. 
There could also be the case that $H$ may become a constant $H_0$ in the infinite future.
However, if $H_0$ is large enough, anything whose binding energy is smaller than a threshold value is also ripped, called a pseudorip~\cite{Frampton:2011aa}.


\section{General anisotropic spacetime}
\label{anisotropic_spacetime}

In this section, we consider a general homogeneous and anisotropic spacetime and classify the future singularities in this spacetime.
In addition to the known finite-time singularities in the FLRW universe,
we show that new kinds of singularities may show up in the anisotropic universe.
We conclude that these singularities require the presence of even slight amount of spacetime anisotropy as a necessary condition.
Regarding the new type of singularities, we investigate the geodesic equation and geodesic deviation equations in such a spacetime.


\subsection{Rotational symmetry breaking}
\label{rotsymbreaking}

The general homogeneous and anisotropic spacetime is given as follows,  
\begin{align}
\label{Eq: metric}
    ds^2 = - dt^2 + \sum_{i,j=1,2,3} g_{ij}(t) dx^i dx^j \, .
\end{align}
The above spacetime is homogeneous because there is a shift symmetry of the spatial coordinates $x^i$, $x^i \to x^i + c^i$ by constants $c^i$. 
Because the spatial part of metric $g_{ij}$ is symmetric under the exchange of the indices $g_{ij}=g_{ji}$, 
we can diagonalize the spatial metric as
\begin{align}
\label{Eq: metric2}
\begin{split}
    \left( g_{ij}\left(t\right) \right) 
    &\equiv 
    \mathcal{O}^\mathrm{T}(t) \left( \tilde{g}_{ij} (t) \right) \mathcal{O}(t)
    \\
    &= \mathcal{O}^\mathrm{T}(t) 
    \left( \begin{array}{ccc} 
    a^{2}(t) & 0 & 0 \\
    0 & b^{2}(t) & 0 \\
    0 & 0 & c^{2}(t) 
    \end{array} \right) 
    \mathcal{O}(t) 
    \, .
\end{split}
\end{align}
Here $\mathcal{O}(t)$ is a $3\times 3$ orthogonal matrix,
and $\mathcal{O}^\mathrm{T}(t)$ is the transpose of $\mathcal{O}(t)$
which satisfies $\mathcal{O}^\mathrm{T}(t) \mathcal{O}(t) = I$ with $3\times 3$ unit matrix $I$. 
$\mathcal{O}(t)$ is time-dependent in general, and if it is a constant matrix, the universe can be regarded as the Bianchi type-I universe, as we will see later.
Note that $\mathcal{O}(t)$ does not mean an actual rotation of space but a rotation of principal axes of the spatial metric represented by a symmetric matrix.

Considering the known results of singularities in the FLRW universe, 
some or all of $a(t)$, $b(t)$, and $c(t)$ may have singularities of type I—V, which is a straightforward generalization of the future singularities in the FLRW universe. 
However, we note that another singularity could be from $\mathcal{O}(t)$. 
Such a singularity is expected to appear under the broken rotational symmetry or the spatial anisotropy.

We now choose the rotational axis of $\mathcal{O}(t)$ near the time $t=t_s$ to be the $x^{3}$-axis, 
which does not generate any loss of generality, 
\begin{align}
\label{Eq: rotation_matrix}
    \mathcal{O}(t)
    = \left( 
        \begin{array}{ccc} 
            \cos\theta(t) & -\sin\theta(t) & 0 \\ 
            \sin\theta(t) & \cos\theta(t) & 0 \\
            0 & 0 & 1 
        \end{array} 
    \right) 
    \, .
\end{align}
As in the cases of type I-IV singularities in the FLRW universe, $\theta(t)$ might have singularities: 
at $t=t_s$,
(1) $\theta(t)$ diverges; 
(2) $\dot\theta(t)$ diverges;
and (3) a higher derivative of $\theta(t)$ diverges. 
We note that the singularity associated with $\theta(t)$ shows up only if $a(t_s)\neq b(t_s)$ because $\mathcal{O}(t)$ becomes irrelevant when $a(t)=b(t)$, 
\begin{align}
\label{Eq: aniso4}
\begin{split}
    &\mathcal{O}^\mathrm{T}(t) 
    \left( 
        \begin{array}{ccc} 
            a^{2}(t) & 0 & 0 \\ 
            0 & a^{2}(t) & 0 \\
            0 & 0 & c^{2}(t) 
        \end{array} 
    \right) 
    \mathcal{O}(t) 
    \\
    &= 
    \left( 
        \begin{array}{ccc} 
            \cos\theta(t) & \sin\theta(t) & 0 \\
            - \sin\theta(t) & \cos\theta(t) & 0 \\
            0 & 0 & 1 
        \end{array} 
    \right) 
    \left( 
        \begin{array}{ccc} 
            a^{2}(t) & 0 & 0 \\
            0 & a^{2}(t) & 0 \\
            0 & 0 & c^{2}(t)
        \end{array} 
    \right) 
    \left( 
        \begin{array}{ccc} 
            \cos\theta(t) & -\sin\theta(t) & 0 \\
            \sin\theta(t) & \cos\theta(t) & 0 \\
            0 & 0 & 1 
        \end{array} 
    \right) 
    \\
    &=
    \left( 
        \begin{array}{ccc} 
            a^{2}(t) & 0 & 0 \\
            0 & a^{2}(t) & 0 \\
            0 & 0 & c^{2}(t)
        \end{array} 
    \right) 
    \, .
\end{split}
\end{align}
Therefore $a(t_s)\neq b(t_s)$ is a necessary condition for the singularity from the rotation $\theta(t)$ along the $x^{3}$-axis. 

For the rotation matrix \eqref{Eq: rotation_matrix}, the spacetime metric \eqref{Eq: metric2} leads to  
\begin{align}
\label{Eq: aniso5}
\begin{split}
    \left( g_{ij}\left(t\right) \right) 
    &= \left( \begin{array}{ccc} 
    a^{2}(t) \cos^2\theta(t) + b^{2}(t) \sin^2\theta(t) & 
    \left[ b^{2}(t) - a^{2}(t) \right] \cos\theta(t) \sin\theta(t) & 0 \\ 
    \left[ b^{2}(t) - a^{2}(t) \right] \cos\theta(t) \sin\theta(t) & 
    a^{2}(t) \sin^2 \theta(t) + b^{2}(t) \cos^2 \theta(t) & 0 \\
    0 & 0 & c^{2}(t) 
    \end{array} \right) 
    \, .
\end{split}
\end{align}
The above expression tells us that if $\theta(t)$ diverges, the metric has violent oscillations, although $\left|g_{ij}\right|$ is finite.  
We will show that there is a curvature singularity even if $\theta(t)$ is finite but $\dot\theta(t)$ diverges. 
Such a singularity occurs when $\theta(t)$ behaves near $t\sim t_s$ as $\theta(t) \sim \theta_0 + \theta_1 \left( t_s - t \right)^\beta$ with constants $\theta_0$, $\theta_1$, and $\beta$ where $0<\beta<1$. 
We briefly comment on the divergence of $\theta$ and its derivatives.
In the spacetime of our interest,
there can be nonzero off-diagonal elements of the spatial metric $g_{ij}(t)$ in Eq.~\eqref{Eq: metric}.
If we assign the scale factors to the diagonal elements of spatial metric $\tilde{g}_{ij}(t)$ after the diagonalization,
the off-diagonal elements in the original spatial metric $g_{ij}(t)$ describes the mixture of the scale factors, as in Eq.~\eqref{Eq: aniso5}.
In this sense, $\theta(t)$ is the time-dependent mixing angle.
Thus, through the diagonalization, $\theta(t)$ corresponds to the off-diagonal elements of $g_{ij}(t)$, $(x,y)$ element in the current setup,
and the divergence of $\theta$ and its derivatives reflects those of such off-diagonal elements in $g_{ij}(t)$.

We compute the Ricci tensor and Ricci scalar in the general homogeneous and anisotropic spacetime (the detailed calculation is summarized in Appendix~\ref{calculationappendix}).
For the metric in Eq.~\eqref{Eq: metric2}, the Ricci tensor is given as follows:
\begin{align}
\label{Eq: ricci_tensor_tt}
\begin{split}
    R_{00}
    &= 
    - \frac{\left( a^{2} - b^{2} \right)^2}{2a^2b^2} \dot\theta^2 
    - \left( 
    \frac{\ddot a}{a} + \frac{\ddot b}{b} + \frac{\ddot c}{c} 
    \right)  
    \, , 
\end{split}
\\
\label{Eq: ricci_tensor_ti}
\begin{split}
    R_{0i} 
    &= R_{i0}
    =0
\end{split}
\\
\label{Eq: ricci_tensor_ij}
\begin{split}
    \left( R_{ij} \right) 
    &\equiv 
    \mathcal{O}^\mathrm{T} ( \tilde{R}_{ij} ) \mathcal{O}
    \\
    &= \mathcal{O}^\mathrm{T}
    \left( 
        \begin{array}{ccc} 
            \tilde{R}_{11} & \tilde{R}_{12} & 0 \\ 
            \tilde{R}_{21} & \tilde{R}_{22} & 0 \\ 
            0 & 0 & \tilde{R}_{33} 
        \end{array} 
    \right) 
    \mathcal{O}
    \, , 
\end{split}
\end{align}
where the components in $\tilde{R}_{ij}$ are defined as
\begin{align}
\label{Eq: ricci_tensor2}
\begin{split}
    \tilde{R}_{11} 
    &=
    \ddot{a}a 
    + \dot{a}a \left(\frac{\dot{b}}{b} +\frac{\dot{c}}{c} \right)
    + \frac{b^4-a^4}{2b^2} \dot{\theta}^2
    \\
    \tilde{R}_{12} 
    &= \tilde{R}_{21}
    \\
    &=
    - \frac{\ddot{\theta}}{2} \left( a^2 - b^2 \right) 
    - \frac{\dot{\theta}}{2}  
    \left[
        \frac{\dot{a}}{a} \left( b^2 + 3a^2\right) 
        -\frac{\dot{b}}{b} \left( a^2 + 3b^2\right)
        +\frac{\dot{c}}{c} \left( a^2 - b^2\right)
    \right] 
    \\
    \tilde{R}_{22} 
    &= 
    \ddot{b}b 
    + \dot{b}b \left(\frac{\dot{a}}{a} +\frac{\dot{c}}{c} \right)
    + \frac{a^4-b^4}{2a^2} \dot{\theta}^2
    \\
    \tilde{R}_{33} 
    &=
    \ddot{c}c
    + \dot{c}c \left( \frac{\dot{a}}{a} +\frac{\dot{b}}{b} \right)
    \, .
\end{split}
\end{align}
By contracting the Ricci tensor with the metric,
the Ricci scalar is given by
\begin{align}
\label{Eq: ricci_scalar}
\begin{split}
    R 
    &=
    \frac{ (a^2 - b^{2})^2 }{4a^{2}b^{2}} {\dot\theta}^2 
    + 2 \left(\frac{\ddot a}{a} + \frac{\ddot b}{b} + \frac{\ddot c}{c} \right)
    + 2 \left( \frac{\dot a \dot b}{ab} + \frac{\dot b \dot c}{bc} + \frac{\dot c \dot a}{ca} \right) 
    \, .
\end{split}
\end{align}
We omitted the variable $t$ in the scale factors and rotation angle above for simplicity.
We note that $\tilde{R}_{33}$ does not have $\theta$ dependence
because we choose the rotation axis as $x^3$ direction in our setup.
One can restore well-known results in FLRW spacetime by taking the limit that $\dot{\theta} = \ddot{\theta} = 0$ and $a = b= c$.

The singularities originated from the rotation angle of spatial metric $\theta$ may show up in the Ricci tensor and Ricci scalar 
if $\dot{\theta}$ or $\ddot{\theta}$ diverge at $t=t_s$.
Notably, these singularities require $a(t_s)\neq b(t_s)$. 
We emphasize that $\theta$ dependence in the curvature tensors always drops if $a(t)=b(t)$, and thus, 
the anisotropy in the scale factors is a necessary condition for the singularity associated with $\theta(t)$.
In other words, if there is even a slight anisotropy in the universe, 
$\theta$ dependence cannot be ignored and potentially causes a new type of singularities.

We consider the case that $\theta(t)$ vanishes at $t=t_s$, while $\dot\theta(t)$ diverges at $t=t_s$. 
In this case, the metric $g_{ij}(t)$ is automatically diagonalized $g_{ij}(t) = \tilde{g}_{ij}(t_s)$ as in Eq.~\eqref{Eq: aniso5}.
However, several components of Ricci tensor $R_{00}$, $R_{11}$, $R_{12}=R_{21}$, $R_{22}$, and Ricci scalar $R$ diverge in general 
when $\dot\theta$ diverges. 
Note that we have chosen the rotational axis as the $x^{3}$ axis near $t=t_s$.
As we will see in the following subsection, in Einstein's gravity, 
the Einstein equation suggests that the energy-momentum tensor must diverge corresponding to divergences of $\dot\theta$ in the Einstein tensor. 
Moreover, off-diagonal components of the Einstein tensor are nonvanishing, which generally requires the anisotropic stress in the energy-momentum tensor.


\subsection{Classification of singularities}
\label{classification}

We can summarize the classification of singularities in terms of the metric components.
First, we consider the singularities related to $a(t)$, $b(t)$, and $c(t)$, which are the eigenvalues of $g_{ij}(t)$. 
When $t \rightarrow t_s$,
\begin{itemize}
    \item[1-1] Type I singularity: 
    Some of $a(t)$, $b(t)$, and $c(t)$ diverge. 
    \item[1-2] Type II singularity: 
    Some of $a(t)$, $b(t)$, and $c(t)$ and the first derivatives of $a(t)$, $b(t)$, and $c(t)$ are finite, but some of the second derivatives diverge. 
    \item[1-3] Type III singularity: 
    Some of $a(t)$, $b(t)$, and $c(t)$ are finite,
    but the first derivatives of $a(t)$, $b(t)$, and $c(t)$ diverge. 
    \item[1-4] Type IV singularity: 
    Some of $a(t)$, $b(t)$, and $c(t)$ and the first and second derivatives of $a(t)$, $b(t)$, and $c(t)$ are finite,
    but some higher derivatives diverge. 
\end{itemize}
Note that the same type of singularity does not need to occur in all directions.
For instance, only $a(t)$ corresponding to $x^1$ direction may have one of the above type I - IV singularities, while the other scale factors do not show singular behaviors.
As another example, the two directions may have singularities, although the remaining direction does not, and these two singularities may be different types from each other. 

As mentioned in the previous subsection, these singularities are generalizations of the future singularities in the FLRW universe with respect to different scale factors assigned to the three spatial directions.
These are not related to the rotation $\theta(t)$,
and assuming the rotation angle is constant $\theta(t)=const.$, 
the spacetime of our interest is reduced to Bianchi type-I universe.
As an illustration, we consider the Einstein equation,
\begin{align}
\label{Eq: einstein_eq}
    G_{\mu\nu} 
    = \kappa^2 T_{\mu\nu} 
    \, ,
\end{align}
where $T_{\mu\nu}$ represents the energy-momentum tensor of the matters.
From Eqs.~\eqref{Eq: metric2} and \eqref{Eq: ricci_tensor_ij}, the Einstein tensor $G_{\mu\nu}$ is given as
\begin{align}
\label{Eq: einstein_tt}
    G_{00} 
    &= 
    - \frac{ (a^2 - b^{2})^2 }{4a^{2}b^{2}} {\dot\theta}^2 
    + \left( \frac{\dot a \dot b}{ab} + \frac{\dot b \dot c}{bc} + \frac{\dot c \dot a}{ca} \right) 
    \, , \\
\label{Eq: einstein_ti}
    G_{0i}
    &= G_{i0} = 0 
    \, , \\
\begin{split}
\label{Eq: einstein_ij}
    G_{ij} 
    &= \mathcal{O}^\mathrm{T} \left(
    \tilde{R}_{ij} - \frac{1}{2} \tilde{g}_{ij} R 
    \right)\mathcal{O} 
    \\
    &\equiv 
    \mathcal{O}^\mathrm{T} ( \tilde{G}_{ij} ) \mathcal{O}
    \\
    &= \mathcal{O}^\mathrm{T}
    \left( 
        \begin{array}{ccc} 
            \tilde{G}_{11} & \tilde{R}_{12} & 0 \\ 
            \tilde{R}_{21} & \tilde{G}_{22} & 0 \\ 
            0 & 0 & \tilde{G}_{33} 
        \end{array} 
    \right) 
    \mathcal{O}
    \, , 
\end{split}
\end{align}
where the diagonal components in $\tilde{G}_{ij}$ are defined as
\begin{align}
\label{Eq: einstein2}
\begin{split}
    \tilde{G}_{11}
    &=
    - \frac{(a^2 - b^2)(3a^{2} + b^{2})}{4b^2} {\dot\theta}^2
    - a^{2} \left(\frac{\ddot b}{b} + \frac{\ddot c}{c} \right)
    - a^{2} \frac{\dot b \dot c}{bc} 
    \, , \\
    \tilde{G}_{22}
    &=
    - \frac{(b^2 - a^2)(a^2 + 3b^2)}{4a^2} {\dot\theta}^2
    -  b^{2}\left(\frac{\ddot a}{a} + \frac{\ddot c}{c} \right)
    -  b^{2} \frac{\dot c \dot a}{ca}
    \, , \\
    \tilde{G}_{33}
    &= 
    - \frac{\left( a^{2} - b^{2} \right)^2 c^{2} }{4a^{2}b^{2}} {\dot\theta}^2
    - c^{2}\left(\frac{\ddot a}{a} + \frac{\ddot b}{b} \right)
    - c^{2} \frac{\dot a \dot b}{ab} 
    \, .
\end{split}
\end{align}
Moreover, the spatial components of the Einstein equation can be simplified by introducing a new definition of the energy-momentum tensor:
\begin{align}
\begin{split}
\label{Eq: einstein_eq2}
    \mathcal{O}^\mathrm{T} ( \tilde{G}_{ij} ) \mathcal{O}
    &=
    \kappa^2 T_{ij}
    \\
    \tilde{G}_{ij}
    &\equiv
    \kappa^2 \tilde{T}_{ij}
    \, ,
\end{split}
\end{align}
where
\begin{align}
    (T_{ij})
    &=
    \mathcal{O}^\mathrm{T}\tilde{T}_{ij} \mathcal{O}
    \, .
\end{align}

Assuming $\theta$ is constant in Eqs.~\eqref{Eq: einstein_tt} and \eqref{Eq: einstein_ij}, 
we find that all the $\theta$-dependent terms vanish, and Eq.~\eqref{Eq: einstein_eq2} leads to the modified Friedmann equations in the Bianchi type-I universe~\cite{Hertzberg:2024uqy,Verma:2023huz}:
\begin{align}
\label{Eq: bianchi1_univ}
\begin{split}
    \kappa^{2} T_{00}
    &= 
    \left( H_{a} H_{b} + H_{b} H_{c} + H_{c}H_{a} \right) 
    \, , \\
    - \frac{\kappa^{2}}{a^2} \tilde{T}_{11}
    &= 
    \left( \dot{H}_{b} + \dot{H}_{c} \right) 
    +\left( H_{b}^2 + H_{c}^2 + H_{b} H_{c} \right)
    \, , \\
    - \frac{\kappa^{2}}{b^2} \tilde{T}_{22}
    &= 
    \left( \dot{H}_{c} + \dot{H}_{a} \right)
    + \left( H_{c}^2 + H_{a}^2 + H_{c}H_{a} \right)
    \, , \\
    - \frac{\kappa^{2}}{c^2} \tilde{T}_{33}
    &= 
    \left( \dot{H}_{a} + \dot{H}_{b} \right)
    + \left( H_{a}^2 + H_{b}^2 + H_{a} H_{b} \right)
    \, .
\end{split}
\end{align}
Here, we defined the Hubble parameter for each direction as
\begin{align}
    H_{a} = \frac{\dot{a}}{a}
    \, , \ 
    H_{b} = \frac{\dot{b}}{b}
    \, , \ 
    H_{c} = \frac{\dot{c}}{c}
    \, .
\end{align}
When we read the energy-momentum tensor as
$T_{00} = \rho$ and $\tilde{T}_{ij} = \mathrm{diag} [ P_1 a^2, P_2 b^2, P_3 c^2]$,
where $\rho$ and $P_i$ are the energy density and the pressure in each direction.
It is apparent that the future singularities in the FLRW universe are generalized into those in the Bianchi Type-I universe.
As in the classification of the future singularity in the FLRW universe,
three different Hubble parameters and their derivatives may show the different types of singularities,
as the corresponding energy density and pressures also diverge.

Second, we focus on the singularities related to the rotation $\theta(t)$ in the orthogonal matrix, which diagonalizes the spatial metric $g_{ij}$. 
For these singularities, the components of the metric are always finite, $\left| g_{ij} \right|<\infty$. 
Near the singularity $t\sim t_s$, we may choose the matrix as in Eq.~\eqref{Eq: rotation_matrix} with any loss of generality and assume $a(t_s)\neq b(t_s)$. 
When $t \rightarrow t_s$,
\begin{itemize}
\item[2-1] Type I$\theta$ singularity: 
$\theta$ diverges, and the metric oscillates very rapidly. 
\item[2-2] Type II$\theta$ singularity: 
$\theta$ and $\dot\theta$ are finite, but $\ddot\theta$ diverges. 
The energy density and the diagonal spatial components of the energy-momentum tensor are finite, 
but the off-diagonal component diverges. 
\item[2-3] Type III$\theta$ singularity: 
$\theta$, $\dot a$ is finite, but $\dot\theta$ and also $\ddot\theta$ diverge. 
The energy density, pressure, and other spatial components of the energy-momentum tensor diverge. 
\item[2-4] Type IV$\theta$ singularity: 
$\theta$, $\dot\theta$, and $\ddot\theta$ are finite, but some higher derivatives of $\theta$ diverge. 
The energy density, pressure, and other spatial components of the energy-momentum tensor are finite.
Their first derivatives are also finite, but the higher derivatives diverge.
\end{itemize}
We note that $\theta(t)$ corresponds to the off-diagonal elements in the original spatial metric $g_{ij}(t)$.
Using a rotation matrix $\mathcal{O}$, we can separate the additional divergence from divergences in the three scale factors.

As before, we consider the Einstein equation as an illustration.
Taking into account the $\theta$-dependence,
we find that the off-diagonal elements in the Einstein tensor \eqref{Eq: einstein_ij} $\tilde{R}_{12} = \tilde{R}_{21}$ do not vanish,
and Eq.~\eqref{Eq: einstein_eq2} leads to the following equations:
\begin{align}
\label{Eq: rotation eom}
\begin{split}
    \kappa^{2} T_{00}
    &= 
    \left( H_{a} H_{b} + H_{b} H_{c} + H_{c}H_{a} \right) 
    - \frac{ (a^2 - b^{2})^2 }{4a^{2}b^{2}} {\dot\theta}^2
    \, , \\
    - \frac{\kappa^{2}}{a^2} \tilde{T}_{11}
    &= 
    \left( \dot{H}_{b} + \dot{H}_{c} \right) 
    + \left( H_{b}^2 + H_{c}^2 + H_{b} H_{c} \right)
    + \frac{(a^2 - b^2)(3a^{2} + b^{2})}{4a^2b^2} {\dot\theta}^2
    \, , \\
    - \kappa^{2} \tilde{T}_{12}
    &=
    \frac{a^2 - b^2}{2} \ddot{\theta}  
    + \frac{1}{2}  
    \left[
        H_{a} \left( b^2 + 3a^2\right) 
        - H_{b} \left( a^2 + 3b^2\right)
        + H_{c} \left( a^2 - b^2\right)
    \right] \dot{\theta}
    \, , \\
    - \frac{\kappa^{2}}{b^2} \tilde{T}_{22}
    &= 
    \left( \dot{H}_{c} + \dot{H}_{a} \right)
    + \left( H_{c}^2 + H_{a}^2 + H_{c}H_{a} \right)
    + \frac{(b^2 - a^2)(a^2 + 3b^2)}{4a^2b^2} {\dot\theta}^2
    \, , \\
    - \frac{\kappa^{2}}{c^2} \tilde{T}_{33}
    &= 
    \left( \dot{H}_{a} + \dot{H}_{b} \right)
    + \left( H_{a}^2 + H_{b}^2 + H_{a} H_{b} \right)
    + \frac{\left( a^{2} - b^{2} \right)^2}{4a^{2}b^{2}} {\dot\theta}^2
    \, .
\end{split}
\end{align}
In addition to the corrections to Eq.~\eqref{Eq: bianchi1_univ}, 
we have an additional equation from the off-diagonal component of the Einstein tensor,
which inevitably introduces the anisotropic stress $\tilde{T}_{12} = \tilde{T}_{21}$.
We can find that in Eq.~\eqref{Eq: rotation eom},
the new types of singularities require anisotropy in the scale factors $a(t)\neq b(t)$ as the necessary condition.
They also indicate that the off-diagonal elements of the energy-momentum tensor, $T_{12}=T_{21}$ in our setup, must have a singularity. 
If $\theta$ behaves as $\theta \sim \theta_0 \left( t_s - t \right)^\beta$ with constants $\theta_0$ and $\beta$ when $t\sim t_s$, 
the type I$\theta$ corresponds to $\beta<0$, the type II$\theta$ to $1<\beta<2$, the type III$\theta$ to $0<\beta<1$, and the type IV$\theta$ to the case that $\beta$ is not an integer and $\beta>2$.


\subsection{Rips and twists}
\label{destruction}

We further investigate the new class of finite-time singularities related to the rotation angle $\theta$ of the spatial metric.
First, we consider what could happen when $\dot\theta$ diverges using the geodesic deviation equation as in Eq.~\eqref{geodesicdev1}. 
Computing the Riemann tensor $R^{i}_{\ ttj}$ in the general homogeneous and anisotropic spacetime, we find
\begin{align}
\begin{split}
    R^i_{\ 00j} 
    &\equiv \mathcal{O}^{T} (\tilde{R}^i_{\ 00j}) \mathcal{O}
    \\
    &=
    \mathcal{O}^{T}
    \left( 
    \begin{array}{ccc} 
        A & D_{ab} & 0 \\ 
        D_{ba} & B & 0 \\ 
        0 & 0 & C 
    \end{array} 
    \right) 
    \mathcal{O}
    \, ,
\end{split} 
\end{align}
and the geodesic deviation equation takes the following form:
\begin{align}
\label{geodesicdev3}
\begin{split}
    \mathcal{O} 
    \left( \begin{array}{c} 
    \frac{D^2 S^1}{d\tau^2} \\ 
    \frac{D^2 S^2}{d\tau^2} \\ 
    \frac{D^2 S^3}{d\tau^2} 
    \end{array} \right) 
    &= 
    \left( 
    \begin{array}{ccc} 
        A & D_{ab} & 0 \\ 
        D_{ba} & B & 0 \\ 
        0 & 0 & C 
    \end{array} 
    \right) 
    \mathcal{O} 
    \left( \begin{array}{c} 
    S^1 \\ 
    S^2 \\ 
    S^3 
    \end{array} \right) 
    \, , 
\end{split}
\end{align}
where
\begin{align}
\begin{split}
    A
    &= 
    (\dot{H}_{a} + H_{a}^{2})  
    - \frac{\dot{\theta}^2}{4} 
    \frac{(a^2-b^2)(a^2+3b^2)}{a^2b^2}
    \\
    D_{ab}
    &=  
    - \frac{\dot{\theta}}{2} 
        \left[
        H_{a} \left( \frac{b^2}{a^2} + 3 \right) 
        - H_{b} \left( 1 + 3 \frac{b^2}{a^2} \right)   
        \right]
    - \frac{\ddot{\theta}}{2} \left( 1 - \frac{b^2}{a^2} \right)
    \\
    D_{ba}
    &= 
    - \frac{\dot{\theta}}{2} 
        \left[
        H_{a} \left( 1 + 3 \frac{a^2}{b^2} \right)
        - H_{b} \left( \frac{a^2}{b^2} + 3 \right)   
        \right]
    - \frac{\ddot{\theta}}{2} \left( \frac{a^2}{b^2} - 1 \right)
    \\
    B
    &=
    (\dot{H}_{b} + H_{b}^{2}) 
    - \frac{\dot{\theta}^2}{4}
    \frac{(b^2-a^2)(b^2+3a^2)}{a^2b^2}
    \\
    C
    &= (\dot{H}_{c} + H_{c}^{2}) 
    \, .
\end{split} 
\end{align}

The off-diagonal components in Eq.~\eqref{geodesicdev3} generate new geodesic deviations proportional to another geodesic deviation perpendicular to the geodesic deviation. 
Especially in the case of type II$\theta$, the diagonal elements are finite, although the off-diagonal elements diverge.
Thus, spacetime could be ripped in analogy to the FLRW case. 
We also consider the analogy to the little rip and pseudorip in the FLRW universe. 
If $\dot\theta$, $\ddot{\theta}$, or both diverge in the infinite future, the spacetime could be ripped finally. 
If $\dot{\theta}$, $\ddot{\theta}$, or both become very large, even constant, any object whose binding energy is below the threshold could be ripped.

Second, we consider what could happen when $\dot\theta$ becomes large, 
using the geodesic equation for the nonrelativistic test particle,
\begin{align}
    \label{geodesic}
    0=\frac{d^2 x^\mu}{ds^2} + \Gamma^\mu_{\rho\sigma} \frac{dx^\rho}{ds} \frac{dx^\sigma}{ds}
    \, .
\end{align}
To investigate effects coming from $\dot\theta$, 
we consider the situation that the divergence from the rotation angle is dominant compared with that from the scale factors in the spatial metric;
that is, we ignore derivatives of the scale factors and assume $\theta \sim 0$ at $t \sim t_s$ as in type III$\theta$ singularity. 
Using the Christoffel symbol in the anisotropic universe (see Appendix~\ref{calculationappendix}),
we find that the spatial components of the geodesic equations lead to
\begin{align}
\label{thetarip}
\begin{split}
    \frac{d^2 x^1}{ds^2} 
    &\sim \dot\theta \left( \frac{b^2}{a^2} - 1 \right) \frac{dx^2}{ds}
    \, , \\
    \frac{d^2 x^2}{ds^2} 
    &\sim \dot\theta \left( 1 - \frac{a^2}{b^2} \right) \frac{dx^1}{ds}
    \, , \\ 
    \frac{d^2 x^3}{ds^2} 
    &\sim 0 
    \, .
\end{split}
\end{align}
Here, we have assumed the nonrelativistic limit $\left| dx^i/ds \right| \ll \left| dx^0/ds \right| \sim 1$.

The terms including $\dot\theta$ generate forces perpendicular to the velocity $dx^i/ds$ of the particle as in the magnetic force. 
These forces may be regarded as the Coriolis force. 
When $\dot\theta$ diverges, any object may be twisted off. 
In this sense, we may call the singularity where $\dot\theta$ diverges in the finite future as the {\it big twist}. 
If $\dot{\theta}$ goes to infinity in the infinite future, we may call this the little twist. 
When $\dot{\theta}$ goes to a very large constant in the infinite future, we may call this phenomenon the pseudotwist.


\section{Reconstruction of models with anisotropic singularity}
\label{reconstruction}

In this section, we consider models that realize curvature singularity by applying a new systematic formulation, so-called reconstruction. 
The reconstruction is the inverse of the standard process where we solve the equations for given models. 
Inversely, we may find a model that realizes the geometry desired from the theoretical and observational viewpoints. 
The reconstruction for cosmology in the FLRW spacetime, Eqs.~\eqref{Eq: FLRWk} and \eqref{Eq: FLRWk0}, has been actively studied for several kinds of modified gravity theories (see the review~\cite{Nojiri:2010wj} and the references therein for the reconstruction, and for modified gravity theories general, see Refs.~\cite{Capozziello:2011et, Nojiri:2017ncd, Arai:2022ilw} for the review). 

Recently, the formulation of the reconstruction for the spherically symmetric spacetime has been investigated, 
using two-scalar fields~\cite{Nojiri:2020blr} and in the scalar-Einstein-Gauss-Bonnet gravity~\cite{Nashed:2021cfs}. 
However, ghosts appear in all the above models, indicating they are physically inconsistent. 
In the classical theory, the kinetic energy of the ghosts is unbounded below, and the system becomes unstable.
In the quantum theory, the ghosts typically generate the negative norm states as in the Fadeev-Popov ghosts in the gauge theories~\cite{Kugo:1979gm}.
The negative norm states generate negative probabilities, which conflicts with the Copenhagen interpretation of the quantum theory.
The ghost can be, however, eliminated by using constraints given by the Lagrange multiplier fields~\cite{Nojiri:2023dvf, Nojiri:2023zlp}. 
We discuss a generalization of the two-scalar model to the model with four scalar fields~\cite{Nashed:2024jqw}. 
This model can reconstruct a model that realizes any given geometry, even if it is time-dependent, not spherically symmetric, and anisotropic, as in Eq.~\eqref{Eq: metric}.


\subsection{Conventional fluid approach}
\label{Sec: fluid approach}

Before we introduce the reconstruction, we consider the effective matter contents that directly reflect the singularities in the Einstein tensor in the framework of Einstein's gravity,
To investigate the new class of singularities generated by $\dot{\theta}$, 
we ignore derivatives of the scale factors and assume $\theta \sim 0$ as done in the previous subsection.
When $\theta \sim 0$, we can drop the rotation matrix $\mathcal{O}$ in Eq.~\eqref{Eq: einstein_eq2},
and the effective energy-momentum tensor of the fluid given by the Einstein tensor is reduced to be 
\begin{align}
\label{Eq: EMT2}
    T_{00} 
    &\sim 
    - \frac{ (a^2 - b^{2})^2 }{4\kappa^2a^{2}b^{2}} {\dot\theta}^2 
    \, , \\ 
    T_{0i} 
    &= T_{i0}=0 
    \, , \\
    \left( T_{ij} \right) 
    &\sim 
    \frac{1}{\kappa^2} 
    {\small
    \left( \begin{array}{ccc}
    - \frac{(a^2 - b^2)(3a^{2} + b^{2})}{4b^2} {\dot\theta}^2 & - \frac{\ddot{\theta}}{2} \left( a^2 - b^2 \right) & 0 \\
    - \frac{\ddot{\theta}}{2} \left( a^2 - b^2 \right) & - \frac{(b^2 - a^2)(a^2 + 3b^2)}{4a^2} {\dot\theta}^2 & 0 \\
    0 & 0 & - \frac{\left( a^{2} - b^{2} \right)^2 c^{2} }{4a^{2}b^{2}} {\dot\theta}^2
    \end{array} \right )
    } 
\end{align}
Although spacetime anisotropy does not allow us to utilize the ordinary perfect fluid description, 
we can define the energy density and pressures $\rho$, $P_{1}$, $P_{2}$, $P_{3}$ as 
\begin{align}
\label{Eq: eff_rho_P}
\begin{split}
    \rho
    &=
    T_{00}  
    \sim 
    - \frac{ (a^2 - b^{2})^2 }{4\kappa^2a^{2}b^{2}} {\dot\theta}^2 
    \, , \\
    P_{1}
    &=
    \frac{T_{11}}{a^2}  
    \sim 
    - \frac{(a^2 - b^2)(3a^{2} + b^{2})}{4\kappa^2a^2b^2} {\dot\theta}^2
    \, , \\
    P_{2}
    &=
    \frac{T_{22}}{b^2}  
    \sim 
    - \frac{(b^2 - a^2)(a^2 + 3b^2)}{4\kappa^2a^2b^2} {\dot\theta}^2
    \, , \\
    P_{3}
    &=
    \frac{T_{33}}{c^2}  
    \sim 
    - \frac{\left( a^{2} - b^{2} \right)^2 }{4\kappa^2a^{2}b^{2}} {\dot\theta}^2
    \, .
\end{split}
\end{align}
Equation~\eqref{Eq: eff_rho_P} shows that
the energy density and three pressures diverge when $\dot{\theta}$ does.
We note that $\rho$ always takes a negative value, which manifestly causes difficulty in introducing the conventional fluid approach at the classical level.
$P_3$ is always negative, while $P_1$ and $P_2$ have opposite signs depending on the scale factor in each direction.

Regarding anisotropic stress, 
if we assume $\theta \sim \theta_0 \left( t_s - t \right)^\beta$ so that $\theta \sim 0$ and $\dot{\theta}$ diverges at $t\sim t_s$,
we find
\begin{align}
\label{Eq: aniso_stress}
    T_{12} &= T_{21} \propto \left|\rho\right|^\frac{\beta-2}{2\left(\beta - 1 \right)} 
    \, .
\end{align}
If we define the power as $\gamma \equiv \frac{\beta-2}{2\left(\beta - 1 \right)}$, 
the type of singularities can be determined by the power $\gamma$:
the type I$\theta$ singularity corresponds to $\frac{1}{2}<\gamma<1$; 
type II$\theta$ to $\gamma<0$;
type III$\theta$ to $\gamma>1$; 
and type IV$\theta$ to $0<\gamma<\frac{1}{2}$ except the points where $\beta$ 
is an integer, that is, $\gamma\neq \frac{n-2}{2 \left( n - 1 \right)}$.
If the universe includes the fluid with the off-diagonal element $T_{12}=T_{21} \propto \left|\rho\right|^\gamma$, there could occur the finite future singularity of 
the type I$\theta$ -- IV$\theta$.

Although the effective fluid cannot be a perfect fluid due to the anisotropy,
we read off EOS in each direction.
Using Eq.~\eqref{Eq: eff_rho_P}, we find
\begin{align}
\label{Eq: EOS}
\begin{split}
    P_{1} &\sim \left( 1 + 2\frac{a^{2} + b^{2}}{a^2 - b^{2}}\right) \rho 
    \, , \\
    P_{2} &\sim \left( 1 - 2\frac{a^{2} + b^{2}}{a^2 - b^{2}}\right) \rho 
    \, , \\
    P_{3} &= \rho
    \, .
\end{split}
\end{align}
Equation~\eqref{Eq: EOS} shows the exotic EOS along $x^1$ and $x^2$ directions depending on the size of the anisotropy.
For $a > b$, the effective EOS paramerter $w>1$ along $x^1$ direction and $w<1$ along $x^2$ direction.
However, the effective fluid shows the stiff EOS $w=1$ along $x^3$ direction regardless of the anisotropy.

Equation~\eqref{Eq: EMT2} suggests that the energy density, pressure, and anisotropic stress of the fluid become smaller if the anisotropy is smaller, $a \sim b$.
Here, we assume a tiny portion of the anisotropic fluid in the present universe, where the background spacetime is almost FLRW Universe,  and we ignore the backreaction of the anisotropic fluid.
For the FLRW metric in \eqref{Eq: FLRWk0}, the Christoffel symbols are given by 
\begin{align}
\label{LC}
\begin{split}
    \Gamma^t_{ij} &= \alpha^2 H \delta_{ij} 
    \, , \\
    \Gamma^i_{tj} &= \Gamma^i_{jt} = H \delta^i_{\ j}
    \, , 
\end{split}
\end{align}
and the other components vanish.
If we impose the conservation law $\nabla^\mu T_{\mu\nu} = 0$ for the anisotropic fluid, 
\begin{align}
\label{cons}
\begin{split}
    0 
    &= \nabla^\mu T_{\mu 0}
    \\
    &= \dot \rho + 3 H \rho + H \left( P_{1} + P_{2} + P_{3} \right) 
    \\
    &= \dot \rho + 6 H \rho 
    \, .
\end{split}
\end{align}
Here we have used \eqref{Eq: EOS} although the EOS could not be valid in the present universe.

On the other hand, the conservation law $\nabla^\mu T_{\mu i} = 0$ is trivial even for the anisotropic fluid in the present model. 
Equation~\eqref{cons} indicates the solution $\rho \propto \alpha^{-6}$, that is, the density decreases by the expansion. 
If the conservation law \eqref{cons} is valid even in the present universe,  the fluid will not dominate in the future. 
The anisotropic fluid cannot describe the future singularity, although it might have been dominant in the early universe and generated the primordial anisotropy. 
In order for the future singularity to show up, when the energy density $\rho$ is small, the EOS \eqref{Eq: EOS} must be changed so that the density increases by the expansion of the universe.


\subsection{Four-scalar reconstruction}

We consider the following model including four scalar fields $\phi^{a}$:
\begin{align}
\label{Eq. acg1}
    S 
    &=
    S_\mathrm{gravity} + S_\phi + S_\lambda 
    \, , \\
\label{Eq. acg2}
    S_\phi 
    &\equiv
    \int d^4x \sqrt{-g} \left(
        \frac{1}{2} \sum_{a,b = 0,1,2,3} A_{ab} \left( \phi \right) 
        g^{\mu\nu} \partial_\mu \phi^{a} \partial_\nu \phi^{b} 
        - V\left( \phi \right) 
    \right)
    \, , \\
\label{Eq. acg3}
    S_\lambda 
    &\equiv
    \int d^4x \sqrt{-g} \sum_{a=0,1,2,3} \lambda^{a} 
    \left( 
        \frac{1}{g^{aa}\left( x = \phi \right)} 
        g^{\mu\nu} \left( x\right) \partial_\mu \phi^{a} \partial_\nu \phi^{a} - 1
    \right) 
    \, .
\end{align}
We use the Roman index ($a, b,\cdots = 0,1,2,3$) for the scalar fields, and as we will see later, it corresponds to the index in the internal space.
$S_\mathrm{gravity}$ represents the action of the arbitrary gravity theory, and the kinetic coefficients $A_{ab} \left(\phi \right)$ and the potential $V\left( \phi \right)$ are functions of the scalar fields $\phi^a$.
In Eq.~\eqref{Eq. acg3}, $\lambda^a$ are Lagrange multiplier fields that lead to constraints,
\begin{align}
\label{Eq: cnstrnt1}
    0 
    = 
    \frac{1}{g^{aa} \left( x = \phi \right)} 
    g^{\mu\nu} \left( x \right) \partial_\mu \phi^{a} \partial_\nu \phi^{a} - 1 
    \, ,
\end{align}
which eliminates ghosts.

By the variation of the action \eqref{Eq. acg1} with respect to the metric $g_{\mu\nu}$, we obtain
\begin{align}
\label{Eq: Eqs1}
\begin{split}
    \mathcal{G}_{\mu\nu} 
    &= 
    \frac{1}{2} g_{\mu\nu} 
    \left(
        \frac{1}{2} \sum_{a, b=0,1,2,3} A_{ab} \left(\phi \right)
        g^{\xi\eta} \partial_\xi \phi^{a} \partial_\eta \phi^{b} 
        - V\left( \phi \right) 
    \right)
    - \frac{1}{2} \sum_{a,b = 0,1,2,3} A_{ab} \left(\phi\right)
    \partial_\mu \phi^{a} \partial_\nu \phi^{b} 
    \\
    & 
    + \frac{1}{2} g_{\mu\nu} \sum_{a=0,1,2,3} \lambda^{a} 
    \left( 
        \frac{1}{g^{aa}\left( x = \phi \right)} 
        g^{\mu\nu} \left( x \right) \partial_\mu \phi^{a} \partial_\nu \phi^{a} - 1
    \right) 
    - \sum_{a=0,1,2,3} \frac{\lambda^{a}}{g^{aa} \left( x = \phi \right)}
        \partial_\mu \phi^{a} \partial_\nu \phi^{a} 
    \\
    &= 
    \frac{1}{2} g_{\mu\nu} 
    \left(
        \frac{1}{2} \sum_{a, b=0,1,2,3} A_{ab} \left(\phi \right)
        g^{\xi\eta} \partial_\xi \phi^{a} \partial_\eta \phi^{b} 
        - V\left( \phi \right) 
    \right)
    \\
    & \qquad 
    - \frac{1}{2} \sum_{a,b=0,1,2,3} A_{ab} \left(\phi \right)
    \partial_\mu \phi^{a} \partial_\nu \phi^{b} 
    - \sum_{a=0,1,2,3} \frac{\lambda^{a}}{g^{aa} \left( x = \phi \right)}
        \partial_\mu \phi^{a} \partial_\nu \phi^{a} 
    \, .
\end{split} 
\end{align}
Here, we used the constraint equations in Eq.~\eqref{Eq: cnstrnt1},
and $\mathcal{G}_{\mu\nu}$ is defined by the variation of the action $S_\mathrm{gravity}$ of the gravity sector: 
\begin{align}
\label{mathcalG}
    \mathcal{G}^{\mu\nu} 
    \equiv
    \frac{1}{\sqrt{-g}} 
    \frac{\delta S_\mathrm{gravity}}{\delta g_{\mu\nu}}
    \, .
\end{align}

If we employ the Einstein-Hilbert action
\begin{align}
\label{Einstein}
    S_\mathrm{gravity} 
    = 
    \frac{1}{2\kappa^2} \int d^4 x \sqrt{-g} R
    \, ,
\end{align}
$\mathcal{G}_{\mu\nu}$ is, of course, given by the Einstein tensor, 
\begin{align}
\label{Eq: GR case}
    \mathcal{G}_{\mu\nu} 
    =
    - \frac{1}{2\kappa^2} G_{\mu\nu}  
    \, .
\end{align}
We can include the contribution of matter by replacing the $\mathcal{G}^{\mu\nu}$ by 
\begin{align}
\label{mathcalGmatter}
\begin{split}
    \mathcal{G}^{\mu\nu} 
    &\equiv 
    \frac{1}{\sqrt{-g}} \frac{\delta S_\mathrm{gravity}}{\delta g_{\mu\nu}}
    + \frac{1}{\sqrt{-g}} \frac{\delta S_\mathrm{matter}}{\delta g_{\mu\nu}}
    \\
    &= 
    \frac{1}{\sqrt{-g}} \frac{\delta S_\mathrm{gravity}}{\delta g_{\mu\nu}}
    + \frac{1}{2} T^{\mu\nu} 
    \, .
\end{split}
\end{align}
Note that the first term is written by the coordinates for a given spacetime metric.
If we find the coordinate dependence of $T^{\mu\nu}$ by solving the conservation law and field equation of the matter, 
the second term and thus the whole $\mathcal{G}^{\mu\nu}$ is written by the coordinates. 
In the case of Einstein's gravity, Eq.~\eqref{Eq: GR case} is rewritten as, 
\begin{align}
\label{Einstein_tensor_matter}
    \mathcal{G}_{\mu\nu} 
    = 
    - \frac{1}{2\kappa^2} G_{\mu\nu} 
    + \frac{1}{2} T_{\mu\nu} 
    \, .
\end{align}

By multiplying Eq.~\eqref{Eq: Eqs1} with $g^{\mu\nu}$, we find
\begin{align}
\label{Eqs2}
    g^{\mu\nu} \mathcal{G}_{\mu\nu} 
    &=
    \frac{1}{2} \sum_{a,b=0,1,2,3} A_{ab} \left(\phi \right) 
    g^{\xi\eta} \partial_\xi \phi^{a} \partial_\eta \phi^{b} 
    - 2 V\left( \phi \right)
    - \sum_{a=0,1,2,3} \lambda^{a} 
    \, ,
\end{align}
where we again used Eq.~\eqref{Eq: cnstrnt1}.
Moreover, substituting Eq.~\eqref{Eqs2} into Eq.~\eqref{Eq: Eqs1}, 
we find
\begin{align}
\label{Eqs3}
    \sum_{a,b=0,1,2,3} A_{ab} \left(\phi\right) \partial_\mu \phi^{a} \partial_\nu \phi^{b}
    &= 
    - 2 \mathcal{G}_{\mu\nu}
    + g_{\mu\nu} \left\{ 
        V\left( \phi \right) 
        + \sum_{a=0,1,2,3} \lambda^{a} 
        + g^{\rho\sigma} \mathcal{G}_{\rho\sigma} 
    \right\} 
    \nonumber \\
    & \qquad 
    - 2 \sum_{a=0,1,2,3} \frac{\lambda^{a}}{g^{aa}\left( x = \phi \right)}
    \partial_\mu \phi^{a} \partial_\nu \phi^{a} 
    \, .
\end{align}
We now identify the four scalar fields as the spacetime coordinates $\phi^{a}=x^a$, 
which is actually consistent with the constraints in Eq.~\eqref{Eq: cnstrnt1}.
And then, Eq.~\eqref{Eqs3} can be rewritten as
\begin{align}
\label{Eqs4}
    A_{\mu\nu} \left(\phi \right) 
    = 
    - 2 \mathcal{G}_{\mu\nu}
    + g_{\mu\nu} 
    \left\{ 
        V\left( \phi \right) 
        + \sum_{a=0,1,2,3} \lambda^{a} 
        + g^{\rho\sigma} \mathcal{G}_{\rho\sigma} 
    \right\}
    - 2 \sum_{a=0,1,2,3} \frac{\lambda^{a}}{g^{aa}\left( x = \phi \right)}
    \delta_{\mu}^{a} \delta_{\nu}^{a} 
    \, .
\end{align}
Moreover, we consider the solution for $\lambda^{a} =0$.
And then, an arbitrary geometry written by $g_{\mu\nu}$ and arbitrary function $ V\left( \phi = x \right)$ can be realized 
by choosing $A_{\mu\nu} \left(\phi \right)$ as
\begin{align}
\label{Eqs5}
    A_{\mu\nu} \left(\phi \right) 
    = - 2 \mathcal{G}_{\mu\nu} \left( x = \phi \right)
    + g_{\mu\nu} \left( x = \phi \right) 
    \left\{ 
        V\left( \phi \right)
        + g^{\rho\sigma} \left( x = \phi \right) \mathcal{G}_{\rho\sigma} \left( x = \phi \right) 
    \right\} 
    \, .
\end{align}
Because the potential $V\left( \phi \right)$ is arbitrary, 
we hereafter choose $V\left( \phi \right)=0$.

We remark several features of $A_{ab}$.
$S_{\phi}$ can be regarded as a nonlinear sigma model whose target-space metric is given by $A_{ab} \left(\phi\right)$ when $V\left( \phi \right)=0$.
A similar structure related to the four scalar fields and internal space can also be found in modified gravity theories~\cite{Hinterbichler:2011tt, Hu:2023gui}.
If $A_{ab}=0$ for a given $a$ and arbitrary $b$ 
and the other nonvanishing components do not depend on $\phi^a$ for the given $a$, 
we may drop the scalar field $\phi^a$. 
For instance, when we consider the spherical symmetry, 
there is no dependence on angular coordinates $\phi^2 = \theta$ and $\phi^3 = \varphi$.
Thus, we can drop two of four scalar fields, and the two-scalar field works for the spherically symmetric spacetime~\cite{Nojiri:2020blr}.

Without $S_\lambda$ in Eq.~\eqref{Eq. acg1}, 
ghosts appear when any eigenvalue of $A_{ab} \left(\phi \right)$ becomes negative.
We now check if the constraints in Eq.~\eqref{Eq: cnstrnt1} derived from $S_\lambda$ can eliminate the ghosts.
For this purpose, we consider the perturbation,
\begin{align}
\label{pert1}
    \phi^a
    = x^a + \delta\phi^a 
    \, .
\end{align}
For the perturbation $\delta\phi^\xi$, 
the constraints in Eq.~\eqref{Eq: cnstrnt1} give
\begin{align}
\label{pert2}
    0 
    = 2 g^{a\nu} \partial_\nu \delta \phi^a - \sum_b \delta \phi^b \partial_b g^{aa}(x)
    \, .
\end{align}
Here, we have not summed the equations with respect to $a$. 
For a spacelike coordinate $x^a$, if we impose $\delta\phi^a=0$ when $\left| x^a \right|\to \infty$, 
and for a timelike coordinate $x^a$, if we impose $\delta\phi^a=0$ as an initial condition, 
we always find $\delta\phi^a=0$.
Therefore, $\delta\phi^a$ does not propagate, and thus the ghosts do not appear.

In the case of Einstein's gravity, Eq.~\eqref{Eqs5} has the following form, 
\begin{align}
\label{Eqs5B}
\begin{split}
    A_{\mu\nu} \left(\phi \right)  
    &= 
    \frac{1}{\kappa^2} G_{\mu\nu} \left( x = \phi \right)
    - \frac{1}{2\kappa^2} g_{\mu\nu} \left( x = \phi \right)  
    g^{\rho\sigma} \left( x = \phi \right) G_{\rho\sigma} \left( x = \phi \right) 
    \\
    &=
    \frac{1}{\kappa^2} R_{\mu\nu} \left( x = \phi \right) 
    \, .
\end{split}
\end{align}
It is now clear that $A_{\mu\nu} \left(\phi \right)$ is given by the Ricci tensor $R_{\mu\nu}$ 
where the coordinates are identified with the scalar fields $x^{\mu} = \phi^{\mu}$. 
Moreover, including matter contents in terms of the energy-momentum tensor as in Eq.~\eqref{Einstein_tensor_matter}, we find
\begin{align}
\label{Eq: Eqs5B_matter}
    A_{\mu\nu} \left(\phi \right) 
    &= \frac{1}{\kappa^2} R_{\mu\nu} \left( x = \phi \right) 
    - T_{\mu\nu} \left( x = \phi \right)  + \frac{1}{2}g_{\mu\nu} \left( x = \phi \right)  T \left( x = \phi \right) 
    \, ,
\end{align}
where $T$ represents the trace of the energy-momentum tensor
\begin{align}
    T
    \equiv g^{\mu\nu} T_{\mu\nu} 
    \, .
\end{align}
Equation~\eqref{Eq: Eqs5B_matter} can be interpreted as $A_{\mu\nu}(\phi)$, which is comprised by the four scalar fields, 
complementing the Einstein equation for any metric $g_{\mu\nu}$ and matter $T_{\mu\nu}$.
Therefore, with an appropriate choice of $A^{\mu\nu}(\phi)$, 
the model described by Eq.~\eqref{Eq. acg1} allows us to reconstruct the gravitational theories that realize the desired geometry.
We note that it is straightforward to extend this reconstruction method to the case in $D$ dimensional spacetime with $D$ scalar fields.


\subsection{Toy model 1: Increasing anisotropy}

We apply the above four-scalar-field model to reconstruct the models that encompass the curvature singularities.
In the homogeneous and anisotropic spacetime described by Eq.~\eqref{Eq: metric}, 
we substitute Eqs.~\eqref{Eq: ricci_tensor_tt} -- \eqref{Eq: ricci_tensor_ij} and Eq.~\eqref{Eq: ricci_scalar} into Eq.~\eqref{Eq: Eqs5B_matter}
\begin{align}
\label{curvatures_aniso3B}
\begin{split}
    A_{00} 
    &= 
    - \frac{1}{\kappa^{2}} \left[ 
        \frac{\left( a^{2} - b^{2} \right)^2}{2a^2b^2} \dot\theta^2 
        + \left( 
            \frac{\ddot a}{a} + \frac{\ddot b}{b} + \frac{\ddot c}{c} 
        \right)  
    \right]_{t=\phi^{(0)}} 
    +  \left[  - T_{00} - \frac{1}{2} T \right]_{t=\phi^{(0)}} 
    \, , \\
    A_{0i}
    &= 
    A_{i0}= - T_{i0} 
    \, , \\
    \left( A_{ij} \right) 
    &= 
    \frac{1}{\kappa^{2}} 
    {\small \left[ 
    \mathcal{O}^\mathrm{T} 
    \left( 
        \begin{array}{ccc} 
            \tilde{R}_{11} & \tilde{R}_{12} & 0 \\ 
            \tilde{R}_{21} & \tilde{R}_{22} & 0 \\ 
            0 & 0 & \tilde{R}_{33} 
        \end{array} 
    \right) 
    \mathcal{O} 
    \right]_{t=\phi^{(0)}} }
    + \left[ - \left( T_{ij} \right) + \frac{1}{2} \left( g_{ij}\right) T  \right]_{t=\phi^{(0)}} \, .
\end{split}
\end{align}

Here, we have denoted $t=x^0$.
We have assumed that the time-dependence of matter, 
and thus $T_{\mu\nu}$ is given by solving the conservation law and field equation of the matter.
We note that $A_{\mu\nu}$ only depends on $\phi^{0}$, $A_{\mu\nu} \left(\phi^0 \right)$, 
because the metric only depends on time coordinate $t=x^{0}$. 
Note that the energy-momentum tensor $T_{\mu\nu}$ represents the ordinary matter contents.
We can utilise the perfect fluid description for $T_{\mu\nu}$, where $T_{i0}=0$, and $A_{\mu\nu}$ compensates the anisotropy.

We reconstruct models realizing the new future singularities discussed in Sec.~\ref{anisotropic_spacetime}, 
demonstrating the four-scalar reconstruction for two different classes of future singularities
by considering the two different situations:
(i) The divergence from the scale factors is dominant compared with that from the rotation angle;
(ii) The divergence from the rotation angle is dominant compared with that from the scale factors.

First, we investigate the case (i) corresponding to Bianchi type-I, where the scale factors may show type I -- IV singularities.
Dropping $\theta$ and its derivatives in Eq.~\eqref{curvatures_aniso3B},
we find that the kinetic coefficient is reduced to
\begin{align}
\label{Eq: toymodel1}
\begin{split}
    A_{00} 
    &= 
    - \frac{1}{\kappa^{2}}
    \left( 
        \frac{\ddot a}{a} + \frac{\ddot b}{b} + \frac{\ddot c}{c} 
    \right)_{t=\phi^{0}} 
    +  \left[  - T_{00} - \frac{1}{2} T \right]_{t=\phi^{0}} 
    \, , \\
    A_{0i}
    &= 
    A_{i0}= 0
    \, , \\
    \left( A_{ij} \right) 
    &= 
    \frac{1}{\kappa^{2}} 
    {\small 
    \left( 
        \begin{array}{ccc} 
            \ddot{a}a + \dot{a}a \left(\frac{\dot{b}}{b} +\frac{\dot{c}}{c} \right) & 0 & 0 \\ 
            0 & \ddot{b}b + \dot{b}b \left(\frac{\dot{a}}{a} +\frac{\dot{c}}{c} \right) & 0 \\ 
            0 & 0 & \ddot{c}c + \dot{c}c \left( \frac{\dot{a}}{a} +\frac{\dot{b}}{b} \right)
        \end{array} 
    \right)_{t=\phi^{0}} }
    + \left[ - \left( T_{ij} \right) + \frac{1}{2} \left( g_{ij}\right) T  \right]_{t=\phi^{0}} 
    \, .
\end{split}
\end{align}
In the above expressions, 
we can choose three scale factors $a(t)$, $b(t)$, $c(t)$ to reconstruct an arbitrary evolution of the background spacetime.
For example, considering a model where the anisotropy vanishes at present $t=t_0$ and grows in the future:
\begin{align}
\label{Eq: scalefactor_ansatz}
\begin{split}
    a(t) 
    &= 
    \alpha(t) \left[ 1 + \tilde{a}(t - t_0)\right]
    \, , \\
    b(t)
    &= 
    \alpha(t) \left[ 1 + \tilde{b}(t - t_0)\right]
    \, , \\
    c(t)
    &= 
    \alpha(t) \left[ 1 + \tilde{c}(t - t_0)\right]
    \, .
\end{split}
\end{align}
Here, $\alpha(t)$ stands for the scale factor in the flat FLRW universe as in Eq.~\eqref{Eq: FLRWk0},
and $\tilde{a}(t)$, $\tilde{b}(t)$, $\tilde{c}(t)$ are increasing functions with respect to $\phi^0$,
which satisfy $\tilde{a}=\tilde{b}=\tilde{c}=0$ at $t=t_0$.
A similar ansatz for the scale factors was discussed in Ref.~\cite{Hertzberg:2024uqy}.

Moreover, if we demand this model mimics the $\Lambda$CDM model in the current universe, 
we include the cosmological constant and dust in the energy-momentum tensor.
Because we are interested in the future singularity, we can assume that the cosmological constant dominates, and then, $T_{\mu\nu}$ is given by
\begin{align}
\label{Eq: CC}
    T_{\mu\nu} = - \frac{\Lambda}{\kappa^2} g_{\mu\nu}
    \, .
\end{align}
We note that the above energy-momentum tensor satisfies the conservation law in the anisotropic universe, and
\begin{align}
\begin{split}
    - T_{00} - \frac{1}{2} T 
    & = \frac{\Lambda}{\kappa^2}
    \, , \\
    - T_{ij} + \frac{1}{2} g_{ij} T 
    & = - \frac{\Lambda}{\kappa^2} g_{ij}
    \, .
\end{split}
\end{align}

Finally, the case (i) can be reconstructed by choosing the following $A_{\mu\nu}$:
\begin{align}
\label{Eq: toymodel12}
\begin{split}
    A_{00} 
    &= 
    - \frac{1}{\kappa^{2}}
    \left( 
        \frac{\ddot a}{a} + \frac{\ddot b}{b} + \frac{\ddot c}{c} 
        - \Lambda
    \right)_{t=\phi^{0}} 
    \, , \\
    A_{0i}
    &= 
    A_{i0}= 0
    \, , \\
    \left( A_{ij} \right) 
    &= 
    \frac{1}{\kappa^{2}} 
    {\small 
    \left( 
        \begin{array}{ccc} 
            \ddot{a}a + \dot{a}a \left(\frac{\dot{b}}{b} +\frac{\dot{c}}{c} \right) - \Lambda a^2& 0 & 0 \\ 
            0 & \ddot{b}b + \dot{b}b \left(\frac{\dot{a}}{a} +\frac{\dot{c}}{c} \right) - \Lambda b^2 & 0 \\ 
            0 & 0 & \ddot{c}c + \dot{c}c \left( \frac{\dot{a}}{a} +\frac{\dot{b}}{b} \right) - \Lambda c^2 
        \end{array} 
    \right)_{t=\phi^{0}} }
    \, .
\end{split}
\end{align}


\subsection{Toy model 2: Rotation singularity}

Second, we consider the case (ii) where the rotation angle $\theta$ may show type I$\theta$-- IV$\theta$ singularities.
Using the setup we used in Sec.~\ref{Sec: fluid approach},
we drop derivatives of the scale factors and assume $\theta \sim \theta_0 \left( t_s - t \right)^\beta$ 
in Eq.~\eqref{curvatures_aniso3B}.
The kinetic coefficient is given by
\begin{align}
\label{Eq: toymode2}
\begin{split}
    A_{00} 
    &= 
    - \frac{1}{\kappa^{2}} \left[ 
        \frac{\left( a^{2} - b^{2} \right)^2}{2a^2b^2} \dot\theta^2 
    \right]_{t=\phi^{0}} 
    +  \left[  - T_{00} - \frac{1}{2} T \right]_{t=\phi^{0}} 
    \, , \\
    A_{0i}
    &= 
    A_{i0}= 0
    \, , \\
    \left( A_{ij} \right) 
    &= 
    \frac{1}{\kappa^{2}} 
    {\small \left[ 
    \left( 
        \begin{array}{ccc} 
            \frac{b^4-a^4}{2b^2} \dot{\theta}^2 &  - \frac{\ddot{\theta}}{2} \left( a^2 - b^2 \right)  & 0 \\ 
            - \frac{\ddot{\theta}}{2} \left( a^2 - b^2 \right)  & \frac{a^4-b^4}{2a^2} \dot{\theta}^2 & 0 \\ 
            0 & 0 & 0 
        \end{array} 
    \right) 
    \right]_{t=\phi^{0}} }
    + \left[ - \left( T_{ij} \right) + \frac{1}{2} \left( g_{ij}\right) T  \right]_{t=\phi^{0}} 
    \, .
\end{split}
\end{align}
Regarding the matter energy-momentum tensor, we can again utilize Eq.~\eqref{Eq: CC}.
Moreover, to mimic the $\Lambda$CDM model, we assume the small but nonzero anisotropy, which is necessary to realize the new types of singularities.
This situation corresponds to $\tilde{a}, \tilde{b}, \tilde{c} \ll 1$ in Eq.~\eqref{Eq: scalefactor_ansatz}.
By the Taylor expansion with respect to $\tilde{a}, \tilde{b}, \tilde{c}$, 
the case (ii) can be reconstructed by the following $A_{\mu\nu}$:
\begin{align} 
\label{Eq: toymode21}
\begin{split}
    A_{00} 
    &= 
    - \frac{1}{\kappa^{2}} 
    \left[ 
        2\left( \tilde{a} - \tilde{b} \right)^2 \dot\theta^2 - \Lambda
    \right]_{t=\phi^{0}} 
    \, , \\
    A_{0i}
    &= 
    A_{i0}= 0
    \, , \\
    \left( A_{ij} \right) 
    &= 
    \frac{\alpha^2(\phi^{0})}{\kappa^{2}} 
    {\small 
    \left( 
        \begin{array}{ccc} 
            2(\tilde{b}-\tilde{a}) \dot{\theta}^2 - \Lambda (1 + 2\tilde{a})
            &  - \ddot{\theta} \left(\tilde{a} - \tilde{b} \right) & 0 \\ 
             - \ddot{\theta} \left(\tilde{a} - \tilde{b} \right)
            & 2(\tilde{a}-\tilde{b}) \dot{\theta}^2 
            -  \Lambda (1 + 2\tilde{b})
            & 0 \\ 
            0 & 0 & - \Lambda (1 + 2\tilde{c})
        \end{array} 
    \right)_{t=\phi^{0}} }
    \, .
\end{split}
\end{align}

We note that the arbitrary divergence of $\theta(t)$ can be reconstructed other than $\theta \sim \theta_0 \left( t_s - t \right)^\beta$.
Moreover, it is optional to include the energy-momentum tensor in this reconstruction method.
When introducing the matter contents, one needs to carefully consider the conservation of the energy-momentum tensor or field equations of matters.
In the above setup, the cosmological constant automatically satisfies the conservation law in our current toy models.


\section{Summary and Discussion}

In this work, we have investigated finite-time singularities in general homogeneous and anisotropic spacetime.
We have observed two classes of singularities.
The first class is associated with the singularities in the scale factors and is the generalization of the well-known finite-time singularities in the FLRW universe.
The second one originates from the spatial anisotropy and rotational symmetry breaking, and the time-dependent rotation angle $\theta (t)$ of the spatial metric may show the new type of singularities.
We have shown that finite anisotropy is the necessary condition for these new singularities, which also introduces the anisotropic stress and off-diagonal elements in the Ricci tensor.
While the divergence of $\theta(t)$ shows violent oscillations in metric, the divergence of its derivatives can occur as $\theta$ vanishes in the future.
Following the finite-time singularities in the FLRW universe, we have categorized the new type of singularities.

We have also considered the physical meanings of divergences in $\theta(t)$
in terms of the geodesic equation and geodesic deviation equation.
In addition to behaviors similar to known results in the FLRW universe, big rip,
we have found a novel singularity named the big twist.
This singularity can be generated by the derivative of $\theta(t)$. 
The big twist shows up in the geodesic equation and is driven by the force perpendicular to the velocity of the test particle,
which is similar to the Coriolis force.
Moreover, we have defined the little twist and pseudotwist based on the behavior of $\dot{\theta}(t)$, which is also analogous to the rip-type singularities in the FLRW universe.

We have finally demonstrated the toy models of finite-time singularities in the homogeneous and anisotropic universe.
The conventional effective matter description in Einstein's gravity, 
where the Einstein tensor directly gives the effective energy-momentum tensor, predicts the exotic equation of state, and it does not work to study the future singularity.
We have developed the novel reconstruction method, the four-scalar reconstruction, and applied it to our consideration.
In the framework of Einstein's gravity, we have reconstructed two models encompassing the two classes of finite-time singularities.
In both models, it is possible to mimic the $\Lambda$CDM model in the current universe, and we can realize the finite-time singularities arising from the scale factor or rotation angle in the spatial metric.

Although we have relied on the reconstruction method in the present work, we can apply our analysis of the finite-time singularities in the homogenous and anisotropic universe to the modified gravity theories beyond Einstein's gravity.
It would be intriguing to study if these singularities, especially newly discovered ones, can be realized in specific models of modified gravity theories.
It would be a realistic extension of existing studies on big rips or other singularities in the FLRW universe in the modified gravity theory.


\acknowledgments

T.K. is supported by the National Key R\&D Program of China (No.~2021YFA0718500)
and by Grant-in-Aid of Hubei Province Natural Science Foundation (No.~2022CFB817).
This work was partially supported by the program Unidad de Excelencia Maria de Maeztu CEX2020-001058-M, Spain (S.D.O).


\appendix

\section{CALCULATION APPENDIX}
\label{calculationappendix}

In this paper, we have defined the Levi-Civita connection, Riemann tensor, Ricci tensor, and Ricci scalar as follows:
\begin{align}
\label{Levi-Civita}
    \Gamma^\sigma_{\mu \nu} 
    &= \frac{1}{2} g^{\sigma \rho} \left( 
    \partial_\mu g_{\rho \nu} + \partial_\nu g_{\rho \mu}- \partial_\rho g_{\mu \nu}
    \right)
    \, , \\
\label{riemann}
    R^\lambda_{\ \mu\rho\nu} 
    & = \Gamma^\lambda_{\mu\nu,\rho} -\Gamma^\lambda_{\mu\rho,\nu} + \Gamma^\eta_{\mu\nu}\Gamma^\lambda_{\rho\eta}
    - \Gamma^\eta_{\mu\rho}\Gamma^\lambda_{\nu\eta} 
    \, , \\ 
\label{riccit}
    R_{\mu\nu} 
    & = R^\rho_{\ \mu\rho\nu} 
    \, , \\
\label{riccis}
    R 
    & = g^{\mu\nu} R_{\mu\nu}
    \, ,
\end{align}


\subsection{Levi-Civita connection}

First, for the metric as in Eq.~\eqref{Eq: metric2}, the Levi-Civita connection \eqref{Levi-Civita} takes the following forms,
\begin{align}
\label{Levi-Civita_aniso}
\begin{split}
    \Gamma^{0}_{00}
    &=
    \Gamma^{0}_{0i}=\Gamma^{0}_{i0} = \Gamma^i_{00}=\Gamma^i_{jk} = 0
    \, , \\
    \Gamma^{0}_{ij} 
    &=
    \frac{1}{2}{\dot g}_{ij}
    \, , \\
    \Gamma^i_{0j}
    &=
    \Gamma^i_{j0}=\frac{1}{2}g^{ik}{\dot g}_{kj}\, .
\end{split}
\end{align}
$\Gamma^{0}_{ij}$ and $\Gamma^i_{0j}$ are written in terms of the rotation matrix $\mathcal{O}$ as follows:
\begin{align}
\label{Levi-Civita_anisoB}
\begin{split}
    \left( \Gamma^t_{ij} \right) 
    &=\frac{1}{2} \left( 
    - \mathcal{O}^\mathrm{T} \dot{\mathcal{O}} \mathcal{O}^\mathrm{T} \tilde{g} \mathcal{O} 
    + \mathcal{O}^\mathrm{T} \dot{\tilde{g}} \mathcal{O} 
    + \mathcal{O}^\mathrm{T} \tilde{g} \dot{\mathcal{O}} 
    \right) 
    \\
    &= \mathcal{O}^\mathrm{T} 
    \left( \begin{array}{ccc} 
    a \dot{a} & \frac{1}{2}\dot\theta \left( b^2 - a^2 \right) & 0 \\
    \frac{1}{2}\dot\theta \left( b^2 -a^2 \right) & b \dot{b} & 0 \\
    0 & 0 &  c\dot{c}
    \end{array} \right) 
    \mathcal{O}
    \, , 
\end{split}
\\
\begin{split}
    \left( \Gamma^i_{tj} \right) 
    &= 
    \frac{1}{2}
    \mathcal{O}^\mathrm{T} \left( \tilde{g} \right)^{-1} \mathcal{O} 
    \left( 
        - \mathcal{O}^\mathrm{T} \dot{\mathcal{O}} \mathcal{O}^\mathrm{T} \tilde{g} \mathcal{O} 
        + \mathcal{O}^\mathrm{T} \dot{\tilde{g}} \mathcal{O} 
        + \mathcal{O}^\mathrm{T} \tilde{g} \dot{\mathcal{O}} 
    \right) 
    \\
    &= 
    \mathcal{O}^\mathrm{T} 
    \left( \begin{array}{ccc} 
    \frac{\dot a}{a} & \frac{1}{2} \dot\theta \left( \frac{b^2}{a^2} - 1 \right) & 0 \\ 
    \frac{1}{2} \dot\theta \left( 1 - \frac{a^2}{b^2} \right) & \frac{\dot b}{b} & 0 \\
    0 & 0 & \frac{\dot c}{c} 
    \end{array} \right) 
    \mathcal{O} 
    \, .
\end{split}
\end{align}
Here, we used
\begin{align}
\label{thetatheta}
\begin{split}
    \dot{\mathcal{O}}^\mathrm{T} 
    &= 
    - \mathcal{O}^\mathrm{T} \dot{\mathcal{O}} \mathcal{O}^\mathrm{T} 
    \\
    \dot{\mathcal{O}}\mathcal{O}^\mathrm{T} 
    &= - \mathcal{O}^\mathrm{T} \dot{\mathcal{O}}
    = \left( \begin{array}{ccc} 
    0 & - \dot\theta & 0 \\
    \dot\theta & 0 & 0 \\
    0 & 0 & 0 
    \end{array} \right)
    \, .
\end{split}
\end{align}


\subsection{Ricci tensor and Ricci scalar}

Second, we compute the Ricci tensor and Ricci scalar:
\begin{align}
\label{curvatures_aniso}
\begin{split}
    R_{00}
    &= - \frac{1}{2} g^{ij}{\ddot g}_{ij} + \frac{1}{2} g^{ij} g^{kl} {\dot g}_{ik} {\dot g}_{jl} 
    - \frac{1}{4} g^{ij} g^{kl} {\dot g}_{ik} {\dot g}_{jl} 
    \\
    &= - \frac{1}{2} g^{ij}{\ddot g}_{ij} + \frac{1}{4} g^{ij} g^{kl} {\dot g}_{ik} {\dot g}_{jl} 
    \, , \\ 
    R_{0i}
    &= R_{i0} = 0 
    \, , \\
    R_{ij}
    &= \frac{1}{2}{\ddot g}_{ij} + \frac{1}{4}{\dot g}_{ij} g^{kl} {\dot g}_{kl} 
    - \frac{1}{2}{\dot g}_{il} g^{lk}{\dot g}_{kj}
    \, , \\ 
    R 
    &= g^{ij}{\ddot g}_{ij} + \frac{1}{4} \left( g^{ij} {\dot g}_{ij} \right)^2 
    - \frac{3}{4} g^{ij} g^{kl} {\dot g}_{ik} {\dot g}_{jl}
    \, .
\end{split}
\end{align}
$R_{00}$, $R_{ij}$, and $R$ in Eq.~\eqref{curvatures_aniso} are given as
\begin{align}
\label{curvatures_aniso2B}
\begin{split}
    R_{00}
    &= \frac{1}{4} \mathrm{tr} \left( 
    - 2 \dot{\mathcal{O}} \mathcal{O}^\mathrm{T} \dot{\mathcal{O}} \mathcal{O}^\mathrm{T} 
    + 2 \dot{\mathcal{O}} \mathcal{O}^\mathrm{T} \dot{\tilde{g}}  \left( \tilde{g} \right)^{-1}  
    + 2 \dot{\mathcal{O}} \mathcal{O}^\mathrm{T} \tilde{g} \dot{\mathcal{O}} \mathcal{O}^\mathrm{T} \left( \tilde{g} \right)^{-1} 
    \right.
    \\
    & \qquad \qquad \qquad
    \left.
    - 2 \left( \tilde{g} \right)^{-1} \dot{\tilde{g}}  \dot{\mathcal{O}} \mathcal{O}^\mathrm{T} 
    + \left( \tilde{g} \right)^{-1} \dot{\tilde{g}} \left( \tilde{g} \right)^{-1} \dot{\tilde{g}}  
    - 2 \ddot{\tilde{g}} \left( \tilde{g} \right)^{-1}
    \right) 
    \, , 
\end{split}
\\
\begin{split}
    R_{ij} 
    &= \frac{1}{4} \left( 
    - \mathcal{O}^\mathrm{T} \dot{\mathcal{O}} \mathcal{O}^\mathrm{T} \tilde{g} \mathcal{O} 
    + \mathcal{O}^\mathrm{T} \dot{\tilde{g}} \mathcal{O} 
    + \mathcal{O}^\mathrm{T} \tilde{g} \dot{\mathcal{O}} 
    \right)_{ij}  
    \mathrm{tr} \left( 
    \left( \tilde{g} \right)^{-1} \dot{\tilde{g}}  \right) 
    \\
    & \quad 
    - \frac{1}{2} \left( 
    \mathcal{O}^\mathrm{T} \ddot{\mathcal{O}} \mathcal{O}^\mathrm{T} \tilde{g} \mathcal{O} 
    - \mathcal{O}^\mathrm{T} \tilde{g} \ddot{\mathcal{O}} 
    - \mathcal{O}^\mathrm{T} \ddot{\tilde{g}} \mathcal{O} 
    - \mathcal{O}^\mathrm{T} \dot{\mathcal{O}} \mathcal{O}^\mathrm{T} \dot{\mathcal{O}} \mathcal{O}^\mathrm{T} \tilde{g} \mathcal{O} 
    \right. \\
    & \qquad \qquad 
    + \mathcal{O}^\mathrm{T} \dot{\mathcal{O}} \mathcal{O}^\mathrm{T} \dot{\tilde{g}}  \mathcal{O} 
    + \mathcal{O}^\mathrm{T} \dot{\mathcal{O}} \mathcal{O}^\mathrm{T} \tilde{g} \dot{\mathcal{O}} 
    - \mathcal{O}^\mathrm{T} \dot{\tilde{g}}  \left( \tilde{g} \right)^{-1} \dot{\mathcal{O}} \mathcal{O}^\mathrm{T} \tilde{g} \mathcal{O} 
    \\
    & \qquad \qquad
    + \mathcal{O}^\mathrm{T} \dot{\tilde{g}}  \left( \tilde{g} \right)^{-1} \dot{\tilde{g}}  \mathcal{O}
    - \mathcal{O}^\mathrm{T} \dot{\tilde{g}}  \dot{\mathcal{O}} 
    - \mathcal{O}^\mathrm{T} \tilde{g} \dot{\mathcal{O}} \mathcal{O}^\mathrm{T} \left( \tilde{g} \right)^{-1} \dot{\mathcal{O}} \mathcal{O}^\mathrm{T} \tilde{g} \mathcal{O}
    \\
    & \qquad \qquad \left. 
    + \mathcal{O}^\mathrm{T} \tilde{g} \dot{\mathcal{O}} \mathcal{O}^\mathrm{T} \left( \tilde{g} \right)^{-1} \dot{\tilde{g}}  \mathcal{O} 
    + \mathcal{O}^\mathrm{T} \tilde{g} \dot{\mathcal{O}} \mathcal{O}^\mathrm{T} \dot{\mathcal{O}} \right)_{ij}
    \, , 
\end{split}
\end{align}
\begin{align}
\begin{split}
    R 
    &= \mathrm{tr} \left( \frac{1}{2} \dot{\mathcal{O}} \mathcal{O}^\mathrm{T} \dot{\mathcal{O}} \mathcal{O}^\mathrm{T} 
    - \frac{1}{2} \dot{\mathcal{O}} \mathcal{O}^\mathrm{T} \dot{\tilde{g}}  \left( \tilde{g} \right)^{-1}  
    - \frac{1}{2} \dot{\mathcal{O}} \mathcal{O}^\mathrm{T} \tilde{g} \dot{\mathcal{O}} \mathcal{O}^\mathrm{T} \left( \tilde{g} \right)^{-1} 
    + \ddot{\tilde{g}}\left( \tilde{g} \right)^{-1}
    + 2 \left( \tilde{g} \right)^{-1} \dot{\tilde{g}}  \dot{\mathcal{O}} \mathcal{O}^\mathrm{T} 
    - \frac{3}{4} \left( \tilde{g} \right)^{-1} \dot{\tilde{g}}  \left( \tilde{g} \right)^{-1} \dot{\tilde{g}}  \right) 
    \\
    & \qquad \qquad 
    + \frac{1}{4} \left( \mathrm{tr} \left( \left( \tilde{g} \right)^{-1} \dot{\tilde{g}}  \right) \right)^2 
    \, ,
\end{split}
\end{align}
We should note that 
\begin{align}
\label{ddottheta}
\begin{split}
    \ddot{\mathcal{O}}\mathcal{O}^\mathrm{T} 
    - \dot{\mathcal{O}}\mathcal{O}^\mathrm{T} \dot{\mathcal{O}}\mathcal{O}^\mathrm{T} 
    &= 
    \left( 
    \begin{array}{ccc} 
    0 & - \ddot\theta & 0 \\ 
    \ddot\theta & 0 & 0 \\ 
    0 & 0 & 0 
    \end{array} 
    \right)
    \ \rightarrow \ 
    \ddot{\mathcal{O}}\mathcal{O}^\mathrm{T} 
    = 
    \left( 
    \begin{array}{ccc} 
    - {\dot\theta}^2 & - \ddot\theta & 0 \\
    \ddot\theta & - {\dot\theta}^2 & 0 \\
    0 & 0 & 0 
    \end{array} 
    \right)
    \, .
\end{split}
\end{align}

By using Eq.~\eqref{thetatheta}, 
$R_{00}$ and $R_{ij}$ of the Ricci tensor are written as follows:
\begin{align}
\begin{split}
    R_{00}
    &= {\dot\theta}^2 \left[ 1  - \frac{1}{2} \left( \frac{b^{2}}{a^{2}} + \frac{a^{2}}{b^{2}} \right) \right]
    - \left( 
    \frac{\ddot a}{a} 
    + \frac{\ddot b}{b} 
    + \frac{\ddot c}{c} \right)  
    \, , 
\end{split}
\\
\begin{split}
    \left( R_{ij} \right) 
    &\equiv 
    \mathcal{O}^\mathrm{T} ( \tilde{R}_{ij} ) \mathcal{O}
    \\
    &= \mathcal{O}^\mathrm{T}
    \left( 
        \begin{array}{ccc} 
            \tilde{R}_{11} & \tilde{R}_{12} & 0 \\ 
            \tilde{R}_{21} & \tilde{R}_{22} & 0 \\ 
            0 & 0 & \tilde{R}_{33} 
        \end{array} 
    \right) 
    \mathcal{O}
    \, , 
\end{split}
\end{align}
where nonzero components in $\tilde{R}_{ij}$ are defined as
\begin{align}
\begin{split}
    \tilde{R}_{11} 
    &=
    \ddot{a}a 
    + \dot{a}a \left(\frac{\dot{b}}{b} +\frac{\dot{c}}{c} \right)
    + \frac{b^4-a^4}{2b^2} \dot{\theta}^2
    \\
    \tilde{R}_{12} 
    &= \tilde{R}_{21}
    \\
    &=
    - \frac{\ddot{\theta}}{2} \left( a^2 - b^2 \right) 
    - \frac{\dot{\theta}}{2}  
    \left[
        \frac{\dot{a}}{a} \left( b^2 + 3a^2\right) 
        -\frac{\dot{b}}{b} \left( a^2 + 3b^2\right)
        +\frac{\dot{c}}{c} \left( a^2 - b^2\right)
    \right] 
    \\
    \tilde{R}_{22} 
    &= 
    \ddot{b}b 
    + \dot{b}b \left(\frac{\dot{a}}{a} +\frac{\dot{c}}{c} \right)
    + \frac{a^4-b^4}{2a^2} \dot{\theta}^2
    \\
    \tilde{R}_{33} 
    &=
    \ddot{c}c
    + \dot{c}c \left( \frac{\dot{a}}{a} +\frac{\dot{b}}{b} \right)
    \, .
\end{split}
\end{align}
And the Ricci scalar $R$ is given as
\begin{align}
\begin{split}
    R 
    &= 
    g^{00} R_{00} + \mathcal{O}^\mathrm{T} \left( \tilde{g}^{ij} R_{ij} \right) \mathcal{O} 
    \\
    &= 
    g^{00} R_{00} + \tilde{g}^{ij} \tilde{R}_{ij}
    \\
    &= 
    \frac{\left( a^{2} - b^{2} \right)^2}{2a^2b^2} \dot\theta^2 
    + 2 
    \left(
    \frac{\ddot a}{a} + \frac{\ddot b}{b} + \frac{\ddot c}{c} 
    \right)
    + 2 
    \left( 
    \frac{\dot a \dot b}{ab} + \frac{\dot b \dot c}{bc} + \frac{\dot c \dot a}{ca} 
    \right) 
    \, .
\end{split}
\end{align}


\subsection{Riemann tensor and geodesic deviation equation}

The spatial components of the geodesic deviation equation as in Eq.~\eqref{geodesicdev} take the following form,
\begin{align}
    \frac{D^2 S^k}{d\tau^2} = R^k_{\ 00j} S^j
    \, .
\end{align}
We compute the Riemann tensor,
\begin{align}
\begin{split}
    R^{0}_{\ i0j} 
    &= 
    \frac{1}{2}{\ddot g}_{ij} + \frac{1}{4} {\dot g}_{ij} g^{lk}{\dot g}_{kl} 
    - \frac{1}{2} {\dot g}_{il} g^{lk}{\dot g}_{kj} 
    \, , 
\end{split}
\end{align}
and thus
\begin{align}
\begin{split}
    R^k_{\ 00j} 
    &= - g^{ki} (g^{00})^{-1} R^0_{\ i0j} 
    = g^{ki} R^0_{\ i0j} 
    \, .
\end{split}
\end{align}
$R^k_{\ 00j}$ is written in terms of the rotation matrix $\mathcal{O}$ as follows:
\begin{align}
    (R^i_{\ 00j}) = \mathcal{O}^{T} (\tilde{R}^i_{\ 00j}) \mathcal{O}
    \, ,
\end{align}
and
\begin{align}
    (\tilde{R}^i_{\ 00j})
    &=
    {\small 
    \left( 
    \begin{array}{ccc} 
        \frac{\ddot{a}}{a} - \frac{\dot{\theta}^2}{4} \frac{(a^2-b^2)(a^2+3b^2)}{a^2b^2} & 
        - \frac{\dot{\theta}}{2} 
        \left[
        \frac{\dot{a}}{a} \left( \frac{b^2}{a^2} + 3 \right) 
        - \frac{\dot{b}}{b} \left( 1 + 3 \frac{b^2}{a^2} \right)   
        \right]
        - \frac{\ddot{\theta}}{2} \left( 1 - \frac{b^2}{a^2} \right) & 
        0 \\ 
        - \frac{\dot{\theta}}{2} 
        \left[
        \frac{\dot{a}}{a} \left( 1 + 3 \frac{a^2}{b^2} \right)
        -\frac{\dot{b}}{b} \left( \frac{a^2}{b^2} + 3 \right)   
        \right]
        - \frac{\ddot{\theta}}{2} \left( \frac{a^2}{b^2} - 1 \right) &
        \frac{\ddot{b}}{b} - \frac{\dot{\theta}^2}{4} \frac{(b^2-a^2)(b^2+3a^2)}{a^2b^2} & 
        0 \\ 
        0 & 0 & \frac{\ddot{c}}{c}  
    \end{array} 
    \right) 
    }
    \, .
\end{align}

\subsection{Einstein tensor}

Finally, we compute the Einstein tensor defined as
\begin{align}
     G_{\mu\nu} = R_{\mu\nu} - \frac{1}{2} g_{\mu\nu} R
     \, .
\end{align}
Using the diagonalized metric and Ricci tensor, 
we can express the spatial components of the Einstein tensor as
\begin{align}
    G_{ij} = \mathcal{O}^\mathrm{T} \left(
    \tilde{R}_{ij} - \frac{1}{2} \tilde{g}_{ij} R 
    \right)\mathcal{O} 
    \, .
\end{align}
Thus, $G_{00}$, $G_{0i}$, and $G_{ij}$ are written as follows:
\begin{align}
\label{Eq: einstein_tensor}
\begin{split}
    G_{00} 
    &= 
    R_{00} + \frac{1}{2} R
    \\
    &= 
    {\dot\theta}^2 \left[ 1  - \frac{1}{2} \left( \frac{b^{2}}{a^{2}} + \frac{a^{2}}{b^{2}} \right) \right]
    - \frac{1}{2}{\dot\theta}^2 \left[
    1 -  \frac{1}{2} \left( 
    \frac{b^{2}}{a^{2}} + \frac{a^{2}}{b^{2}} \right) 
    \right]
    + \left( \frac{\dot a \dot b}{ab} + \frac{\dot b \dot c}{bc} + \frac{\dot c \dot a}{ca} \right) 
    \\
    &= 
    - \frac{ (a^2 - b^{2})^2 }{4a^{2}b^{2}} {\dot\theta}^2 
    + \left( \frac{\dot a \dot b}{ab} + \frac{\dot b \dot c}{bc} + \frac{\dot c \dot a}{ca} \right) 
    \, , 
\end{split}
\\
\begin{split}
    G_{0i}
    &= G_{i0} = 0 
    \, , 
\end{split}
\\
\begin{split}
    \left( G_{ij} \right) 
    &= \mathcal{O}^\mathrm{T} \left \{ 
    \left( 
        \begin{array}{ccc} 
            \tilde{R}_{11} & \tilde{R}_{12} & 0 \\ 
            \tilde{R}_{21} & \tilde{R}_{22} & 0 \\ 
            0 & 0 & \tilde{R}_{33} 
        \end{array} 
    \right) 
    - \frac{1}{2} R
    \left( \begin{array}{ccc} 
    a^{2} & 0 & 0 \\
    0 & b^{2} & 0 \\
    0 & 0 & c^{2}
    \end{array} \right)
    \right\} 
    \mathcal{O} 
    \\
    &= \mathcal{O}^\mathrm{T}
    \left( 
        \begin{array}{ccc} 
            \tilde{G}_{11} & \tilde{R}_{12} & 0 \\ 
            \tilde{R}_{21} & \tilde{G}_{22} & 0 \\ 
            0 & 0 & \tilde{G}_{33} 
        \end{array} 
    \right) 
    \mathcal{O}
    \, ,
\end{split}
\end{align}
where the diagonal components in $\tilde{G}_{ij}$ are defined as
\begin{align}
\begin{split}
    \tilde{G}_{11}
    &=
    - \frac{(a^2 - b^2)(3a^{2} + b^{2})}{4b^2} {\dot\theta}^2
    - a^{2} \left(\frac{\ddot b}{b} + \frac{\ddot c}{c} \right)
    - a^{2} \frac{\dot b \dot c}{bc} 
    \, , \\
    \tilde{G}_{22}
    &=
    - \frac{(b^2 - a^2)(a^2 + 3b^2)}{4a^2} {\dot\theta}^2
    -  b^{2}\left(\frac{\ddot a}{a} + \frac{\ddot c}{c} \right)
    -  b^{2} \frac{\dot c \dot a}{ca}
    \, , \\
    \tilde{G}_{33}
    &= 
    - \frac{\left( a^{2} - b^{2} \right)^2 c^{2} }{4a^{2}b^{2}} {\dot\theta}^2
    - c^{2}\left(\frac{\ddot a}{a} + \frac{\ddot b}{b} \right)
    - c^{2} \frac{\dot a \dot b}{ab} 
    \, .
\end{split}
\end{align}

\bibliography{reference}

\end{document}